\documentclass[runningheads]{svmult}

\usepackage{makeidx}   
\usepackage{graphicx}  
\usepackage{subeqnar}  
\usepackage{multicol}  
\usepackage{physprbb}  
\makeindex             

\newcommand{\msun}{M_{\odot}}

\newcommand{\beqn}{\begin{eqnarray}}
\newcommand{\eeqn}{\end{eqnarray}}

\begin{document}
\title*{Signal of Quark Deconfinement in 
Millisecond\protect\newline Pulsars
and Reconfinement in Accreting X-ray Neutron Stars}
\toctitle{Signal of Quark Deconfinement in 
Millisecond\protect\newline Pulsars
and Reconfinement in Accreting X-ray Neutron Stars}
%
%
\titlerunning{Signal of Quark Deconfinement}
%
\author{Norman~K.~Glendenning\inst{1}
\and Fridolin Weber\inst{2}}
%
\authorrunning{Norman~K.~Glendenning \& Fridolin Weber}
%
%
\institute{Nuclear Science Division, \&
Institute for Nuclear and Particle Astrophysics,\\  Lawrence Berkeley 
National Laboratory,\\
University of California, Berkeley, CA 94720
\and 
University of Notre Dame\\
Department of Physics\\
225 Nieuwland Science Hall\\
Notre Dame, IN 46556-5670}

\maketitle              

\begin{abstract}
Theoretically, the phase transition between the confined
and deconfined phases of quarks can have a remarkable effect
on the spin\index{Neutron star!spin properties}
 properties of millisecond pulsars and on the spin distribution
of the population of
x-ray\index{Binaries!x-ray} neutron stars in low-mass binaries. In the latter class of stars, the effect has already been observed---a strong clustering
in the population in a narrow band of spins. The observed 
clustering cannot presently
 be uniquely assigned to the phase transition as cause. However,
there is another possible signal---not so far observed---in 
millisecond pulsars that we also discuss which would have the
same origin, and whose discovery would tend to confirm the
interpretation in terms of a phase transition in the stellar core.
\end{abstract}
\section{Motivation}

One of the great fascinations of neutron stars
is the deep interior where the density is a few times 
larger than the density of normal nuclei. There, in matter
 inaccessible to us other than fleetingly in relativistic 
collisions, 
unfamiliar states  may exist. The most exotic of these is, 
of\index{Quark matter}\index{Deconfined hadronic matter|see Quark
matter} 
course, quark matter---the deconfined phase of hadronic matter. 
It is quite plausible that
ordinary canonical\index{Pulsar!canonical} pulsars---those like the
Crab\index{Crab pulsar}, 
and more slowly rotating ones---have a quark matter core, or 
at least a mixed phase of quark and confined hadronic matter. 
If one were able to close pack nucleons to a distance such that they
were touching at  a radius of 0.5 fm, the density would be a mere  
$ 1/[(2r)^3\rho_0] =6.5 $ larger than normal nuclear density. 
If nucleons were close packed to their rms
charge radius of 0.8 fm, the density would be 1.6 times nuclear 
density. Of course, fermions cannot be close packed according to the
exclusion principle since their localization would give them 
enormous 
uncertainty in their momentum. They would be torn apart before they
could be squeezed so much.  This simple, and perhaps oversimplified 
argument, makes it plausible  that ordinary (slowly rotating pulsars) 
have a quark matter core essentially from birth.

By comparison, millisecond\index{Pulsar!millisecond}
 (ms) pulsars are centrifugally flattened in the 
equatorial plane and the density is diluted in the
interior. We shall suppose\index{Phase transition!in neutron star}
that the critical phase transition density lies between the 
diluted density of millisecond pulsars and the high density at the
center of canonical pulsars. Then as a millisecond
neutron star spins down because it is radiating angular momentum 
in a broad band of electromagnetic frequencies as well as in a wind
of particle-antiparticle pairs, or as a canonical neutron star at some 
stage begins accreting matter from a companion and is spun up, 
a change in phase of matter in the inner part of the star will occur.

Whatever the high density phase, and we assume here it is the
quark deconfined one, it is softer than the normal phase. Otherwise
no\index{Equation of state!softening}
 phase transition. Consequently, a change in the phase of matter
will be accompanied by a change in the density distribution in the 
star. In the case of spindown of an 
isolated ms pulsar,  the weight of the surrounding part of the star will 
squeeze the softer high density phase that is forming  in the core.  
Conversely, an accreting neutron star that is being spun up, will spin 
out the  already present quark phase\index{Quark matter}. In either case, the star's  
moment of inertia\index{Neutron star!moment of inertia}
 will change, and therefore 
its spin rate will change to accommodate the conservation of angular 
momentum. Timing of pulsar and x-ray neutron star rotation is
a relatively easy observation, and the effect of a phase change 
in the deep interior of neutron stars on timing or  frequency 
distribution is what we study here. 

The deconfinement\index{Phase transition!deconfinement}
 or reconfinement of\index{Phase transition!in neutron star} 
quark matter\index{Quark matter} in a rotating star 
is a very slow process because it is governed by the rate of change of 
period, which is very small. This is an advantage. If the processes 
were fast,  we would not likely witness the epoch of phase change, 
which is long,  but nevertheless
short (but not too short) compared to the 
timescales of spindown of ms pulsars\index{Pulsar!millisecond}
 or of spinup of accreting\index{Neutron star!accretion} 
neutron\index{Neutron star!in low-mass x-ray binary} stars in low-mass binaries\index{Binaries!x-ray}. What we may see in the 
case\index{Accretion!onto neutron star} of isolated ms pulsars is an occasional one that is spinning up, 
even though losing angular momentum to radiation.  That would be a 
spectacular signal. What we may see in the case of accreting 
x-ray\index{Neutron star!x-ray accreter}\index{Neutron star!in 
low-mass x-ray binary}\index{X-ray!accreter} 
neutron stars in low-mass\index{Binaries!x-ray} 
binaries is an unusual number of them falling within 
a small spin-frequency range---the range that corresponds to the spinout of the\index{Quark matter} 
quark matter phase. That also is an easily observable signal.   
For stars of the same mass, the spontaneous spinup of
isolated pulsars should occur at about the same spin frequency 
as the
clustering in the population of neutron star accretors.

\begin{figure}[t] 
\begin{center} 
\includegraphics[width=.6\textwidth]{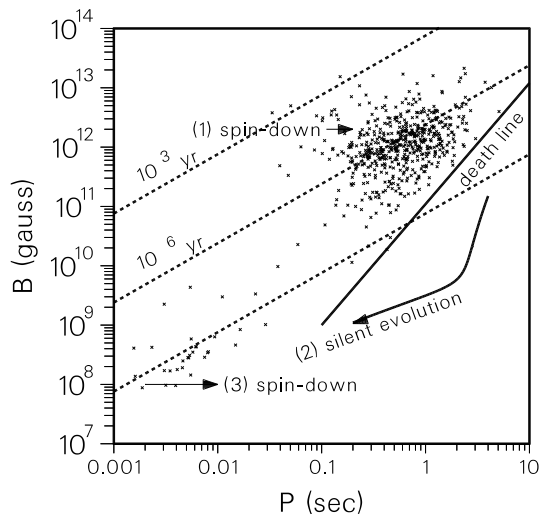} 
\end{center} 
\caption[]{Magnetic field of radio pulsars as a function of rotation period. 
Death line shows a combination of the two properties beyond  which the 
beaming mechanism apparently fails. Canonical pulsars are  designated as (1), 
millisecond pulsars as (3) and  a schematic path for x-ray accreting neutron 
stars as (2). From Ref.\ \protect\cite{books}.}. 
\label{pulsars123} 
\end{figure}

To place the canonical pulsars, ms pulsars, and x-ray accreting 
neutron\index{Canonical pulsar}\index{Millisecond pulsar} stars in context, we refer to Fig.\ \ref{pulsars123}. 
Canonical\index{X-ray!neutron star} 
pulsars have  large surface magnetic fields, of the order 
of $10^{12}{\rm~to~}10^{13}$ G, and relatively long periods with an 
average of about 0.7 s. The ms pulsars have low fields, of the order 
of $10^8{\rm~to~}10^9$ G, and periods ranging from 1 to 10 ms. 
The x-ray stars that are accreting matter from a  low-mass 
non-degenerate
companion  are believed to be the link  
\cite{alpar82:a,heuvel91:a,klis98:b,chakrabarty98:a} 
between the populations of canonical and ms pulsars---their path is 
indicated  schematically in the figure. (Of course, defining ms pulsars to be 
those lying in the range of periods  1 to 10 ms is arbitrary.)

At present, fewer ms pulsars have been discovered than
canonical pulsars. This is likely to
be a selection effect. 
It is much more difficult to detect ms pulsars because there is radio 
noise in all directions (and therefore a possible signal)
and because of the\index{Interstellar medium!dispersion of radio signal} 
dispersion\index{Dispersion by interstellar electrons} by the interstellar electrons (from ionized hydrogen)
 of any pulsed signal which
might be present in the direction the telescope is pointing. 
If present the signal 
is weak and contains a band of radio frequencies. Because of the 
different time-delay of each frequency, the frequencies are 
placed into a number of bins. For ms pulsars, the
time-delay across the range of frequencies in a pulse is 
greater than the interval between pulses. Detection therefore depends
on a tedious analysis that corrects the time-delay in each 
frequency bin
for an assumed density of intervening interstellar gas;
this is repeated for a succession of assumed densities. If the 
process converges to a periodic pulse, a pulsar has been discovered. 
Otherwise, the radio telescopes are pointed in a new direction.
Systematically covering the sky is obviously very costly and
time consuming, and moreover, the present instrumentation
is unable to detect pulsars with periods less than 1 millisecond.

\section{Spontaneous Spinup of Millisecond Pulsar}

We discuss here the effect that a change of phase in the core of a
neutron star can have on the rotational properties of ms pulsars
\cite{glen97:a,glen97:e,heiselberg98:a,blaschke99:a}. 
The analysis begins with the energy-loss equation
for a rotating magnetized star whose magnetic axis is tilted with
respect to the rotation axis. It is of historical interest
to note that even before pulsars were discovered
and soon identified with hypothetical neutron stars, Pacini had
postulated the existence of a highly magnetized rotating neutron star
inside the Crab\index{Crab pulsar!power output}
 nebula as the energy source of the 
nebula and inferred some of the star's properties
\cite{pacini67:a}. It was already known before 1967 that
the nebula, formed in a supernova in 1054, is being  accelerated
and illuminated by a power source amounting to $\sim 4 \times
10^{38}$ erg/s. This compares with a solar luminosity of
$L_{\odot} \sim 4 \times 10^{33}$ erg/s. The power output of the
rotating Crab pulsar equals that of 100,000 suns. Equating the 
power input to the nebula with the power radiated by 
 a rotating magnetic 
dipole, 
\begin{eqnarray}
4\times 10^{38} {\rm~ergs/s} =-\frac{dE}{dt} =
- \frac{d}{dt}\Bigl(\frac{1}{2} I \Omega^2 \Bigr)
= \frac{2}{3} R^6 B^2 \Omega^4 \sin^2 \alpha \,,
\label{dipole}
\end{eqnarray}
and knowing\index{Crab pulsar!properties}
 the period and rate of change of period of the pulsar,
\begin{eqnarray}
P\sim \frac{1}{30}~{\rm s},~~~~~
\dot{P} \sim 4 \times 10^{-13}~{\rm s/s}\,,
\end{eqnarray}
the moment of inertia, surface magnetic field strength,
 and rotational
energy of the Crab pulsar can be inferred as 
\begin{eqnarray}
\begin{array}{l}
I \sim 9
\times 10^{44}~ {\rm g~cm^2} \sim 70 {\rm~km^3}\,, \\[.5ex]
B \sim 4 \times 10^{12} ~ {\rm gauss}\,, \\[.5ex]
E_{{\rm rot}}
\sim \frac{1}{2} I \Omega^2 \sim 2 \times 10^{49}~ {\rm ergs }
\sim 10^{55}{\rm~ MeV}\,.
\end{array} \label{crabprop}
\end{eqnarray}
For these estimates,
we have assumed that $\sin \alpha =1$, where $\alpha$ is the
angle between magnetic and rotational axis. 
(Gravitational units $G=c=1$ are used frequently. For convenient
conversion formula to other units see ch. 3 of Ref.\ \cite{books}.)

Returning to the effect that a phase change can have on the
rotational properties of a pulsar\index{Neutron star!moment of inertia} through changes in the moment
of inertia\index{Phase transition!and moment of inertia} induced by a phase transition, 
we rewrite Eq.\ (\ref{dipole}) in greater generality:
\begin{eqnarray}
\dot{E}= \frac{d}{dt}\biggl(\frac{1}{2} I \Omega^2 \biggr)=-C\Omega^{n+1}\,.
\label{loss}
\end{eqnarray}
Here we have written for convenience
\begin{eqnarray}
C=(2/3) R^6 B^2 \sin^2 \alpha\, .
\label{C}
\end{eqnarray}
We find the deceleration equation
\begin{eqnarray}
\dot{\Omega}= -\frac{C}{I} \Omega^n
\biggl(1 + \frac{I^{\prime}\Omega}{2I}\biggr)^{-1}
\, .
\label{braking2}
\end{eqnarray}
In work previous to  ours, $I^\prime\equiv dI/d\Omega$ was assumed
to vanish. This would be a good approximation for canonical pulsars
but not for millisecond pulsars. 

The angular momentum of a rotating star
 in General Relativity can be obtained numerically
as a solution of Einstein's equations or as a very
complicated algebraic
expression in a perturbative expansion. The moment of inertia
is then\index{Neutron star!moment of inertia}
 obtained as $I=J/\Omega$. The complication arises in two
ways. First,
a  rotating star sets the local inertial frames into rotation.
This is referred to as frame dragging. Second, the structure of
the rotating star depends on the frame dragging frequency, 
$\omega$, which
is position dependent, and on the spacetime metric, which also
depends on $\omega$. Algebraic expressions were obtained 
in Refs.\ \cite{glen92:b,blaschke99:a}.

\begin{figure}[t]
\vspace{-.24in}
\begin{center}
\includegraphics[width=.5\textwidth]{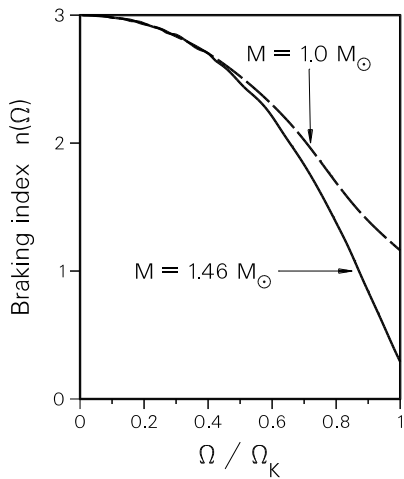}
\end{center}
\caption[]{Apparent braking index  for pure dipole radiation ($n=3$) 
as function of rotational\index{Braking index!as function of stellar spin}
frequency for stars of two different masses containing hyperons 
(but no first order phase transition).
(The mass indicated is for slow rotation.)
\label{hyp_n}
}.
\end{figure}

The index $n$ in equation Eq.\ (\ref{loss}) equals three for magnetic
dipole radiation. In principle, it can be measured in terms of the
frequency $\Omega$ and its first two time derivatives. But the
dependence of $I$ on $\Omega$ introduces a correction so that
the dimensionless {\sl measurable} braking index\index{Braking index}
 is given by\index{Neutron star!braking index|see Braking index}
\begin{eqnarray}
n(\Omega) = \frac{\Omega \ddot{\Omega} }{\dot{\Omega}^2}
= n
- \frac{ 3 I^\prime \Omega +I^{\prime \prime} \Omega^2 }
{2I + I^\prime \Omega} \,.
\label{vindex}
\end{eqnarray}
So the measurable braking index will differ from 3 for ms pulsars,
if for no other reason than that the centrifugal deformation
of the star relaxes as the star spins down. Such a dependence on 
$\Omega$ is illustrated in Fig.\ \ref{hyp_n}

 It is especially
noteworthy that the {\sl magnitude} of the signal---the deviation
of $n(\Omega)$ from $n$---depends on $dI/d\Omega$ which is large, especially
for millisecond pulsars,
but the {\sl duration} of the signal depends on $d\Omega/dt$ which is small.
Therefore,
the variation
of the braking index
 over time is very slow, the time-scale being astronomical.
So what one would observe over any
observational era is a constant braking index $n(\Omega)$
that\index{Braking index!as function of stellar spin}
is
less (if no phase transition)
 than the $n$ characteristic of the energy-loss mechanism 
Eq.\ (\ref{loss}) even if the
radiation were a pure multipole.
We will see that the effect of a phase transition on the 
measurable braking
index can be (but not necessarily)
 much more dramatic. (We distinguish between the constant
$n=3$ of the energy-loss equation and the measurable quantity
of Eq.\ (\ref{vindex}).) 

\begin{figure}[tbh]
\vspace{-.24in}
\begin{center}
\leavevmode
\centerline{ \hbox{
\includegraphics[width=.5\textwidth]{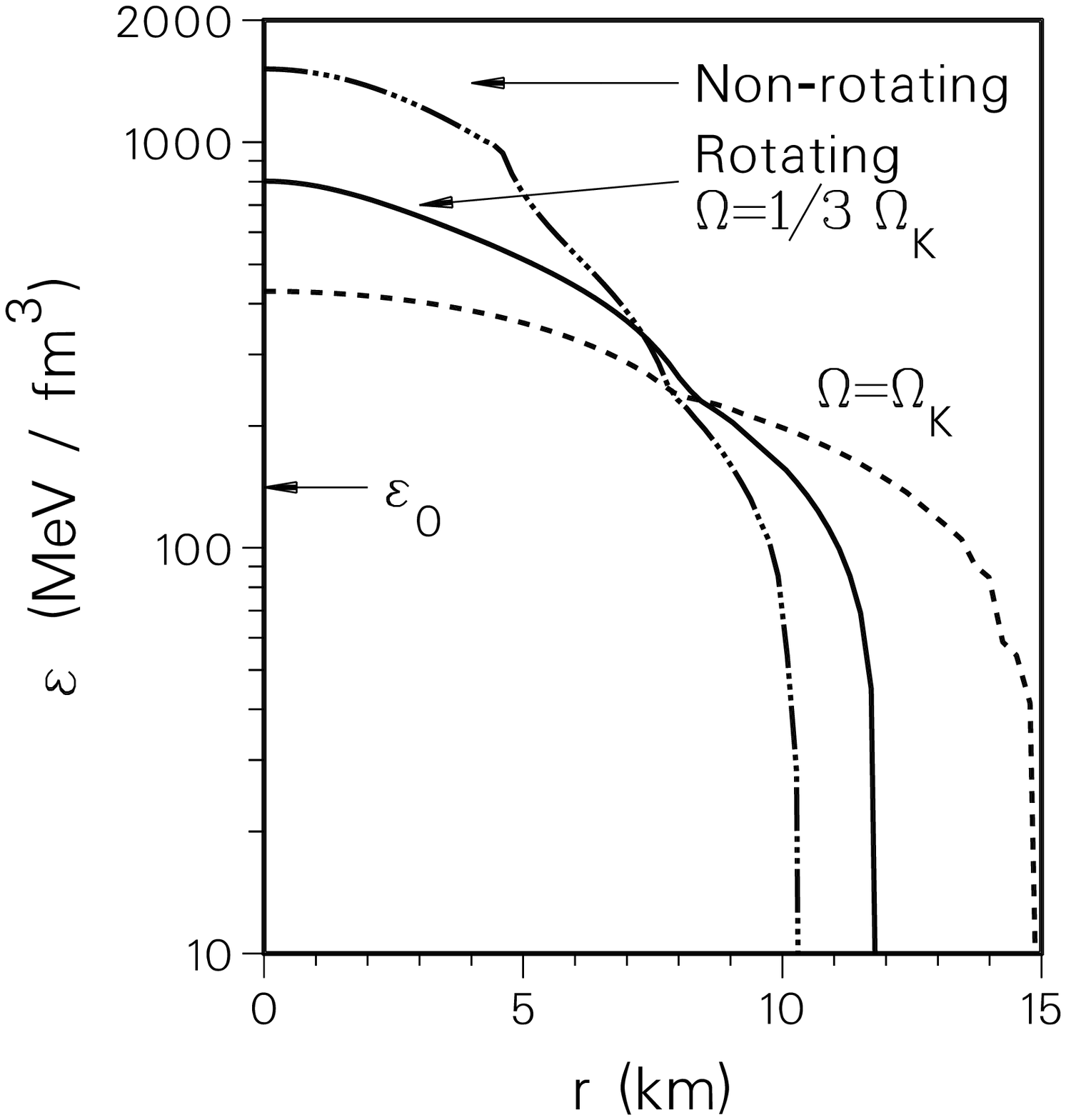}
\includegraphics[width=.5\textwidth]{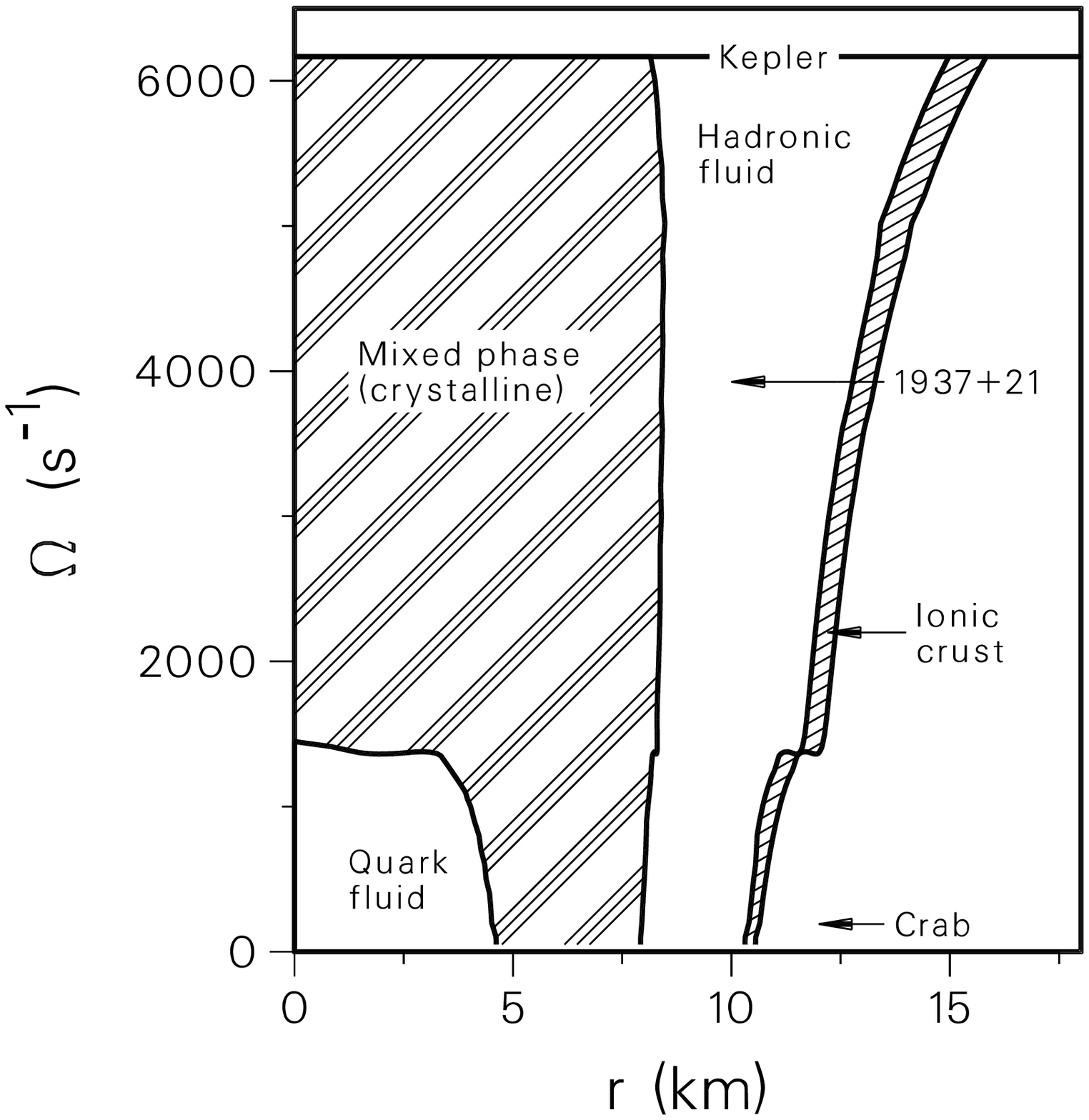}
}}
\begin{flushleft}
\parbox[t]{1\textwidth} { \caption { Mass profiles as a function 
of equatorial 
radius
of a star rotating at three different frequencies,
as marked. At low frequency the star is very dense in its core, having
a 4 km central region\index{Quark matter}
 of highly compressible pure quark matter.
At intermediate frequency, the pure quark matter phase is absent and
the central 8 km is occupied\index{Quark matter!mixed phase} 
by the mixed phase. At higher 
frequency (nearer $\Omega_K$) the star is relatively
dilute in the center and centrifugally stretched.  Inflections at
$\epsilon=220$ and $950$ are the boundaries of the mixed phase.
\label{prof_k300b180}
}} 
\parbox[t]{1\textwidth} { \caption { Radial boundaries at various 
rotational frequencies
separating  (1) pure quark matter,  (2)
mixed phase, (3) pure hadronic phase, (4) ionic crust of neutron 
rich nuclei and surface of star. The pure\index{Quark matter}
quark phase appears only when the  frequency is below $\Omega
\sim 1370$ rad/s. Note the decreasing radius as the frequency falls.
The frequencies of two pulsars, the Crab and
PSR 1937+21 are marked for reference.
\label{omega_r_k300B180}
}}
\end{flushleft}
\end{center}
\end{figure}

We turn now to our original postulate---that canonical pulsars
have a quark matter\index{Quark matter}
 phase in their central region but that 
 ms pulsars  which are
centrifugally diluted, do not. In the slow course of spindown the
central density will rise however. When the critical density is 
reached,
stiff  nuclear matter will be  replaced slowly in the core  by
highly compressible quark matter. The overlaying layers of nuclear matter 
weigh down on the core
and compress it. Its density rises. The star shrinks---mass
is redistributed with growing concentration at the center.
The by-now more massive central region gravitationally
compresses the outer nuclear matter even further, amplifying
the effect. The density profile for a star at three angular
velocities,  (1) the limiting Kepler
angular  velocity at which the star is stretched
in the equatorial plane and its density is
centrally diluted,  (2) an
intermediate angular velocity, and  (3) a non-rotating star,  are
shown in Fig.\  \ref{prof_k300b180}.
We see that the central density rises with
decreasing angular velocity by a factor of three and the
equatorial radius decreases by 30 percent.
In contrast, for a model for
which the phase transition did not take place, the central
density would change by only a few percent
\cite{weber90:d}.
The phase boundaries
are shown in Fig.\ \ref{omega_r_k300B180} from the highest
rotational frequency to zero rotation.

The redistribution of mass and shrinkage of the star change
its moment of inertia\index{Neutron star!moment of inertia}
 and hence the characteristics of its
spin behavior. The star must spin up to conserve angular momentum
which is being carried off only  slowly by the weak electromagnetic
dipole radiation. The star  behaves like an ice skater
who goes into a spin with arms outstretched, is slowly spun
down by friction, temporarily spins up by pulling the arms
inward, after which friction takes over again.

\begin{figure}[ht]
\vspace{-.24in}
\begin{center}
\leavevmode
\centerline{ \hbox{
\includegraphics[width=.5\textwidth]{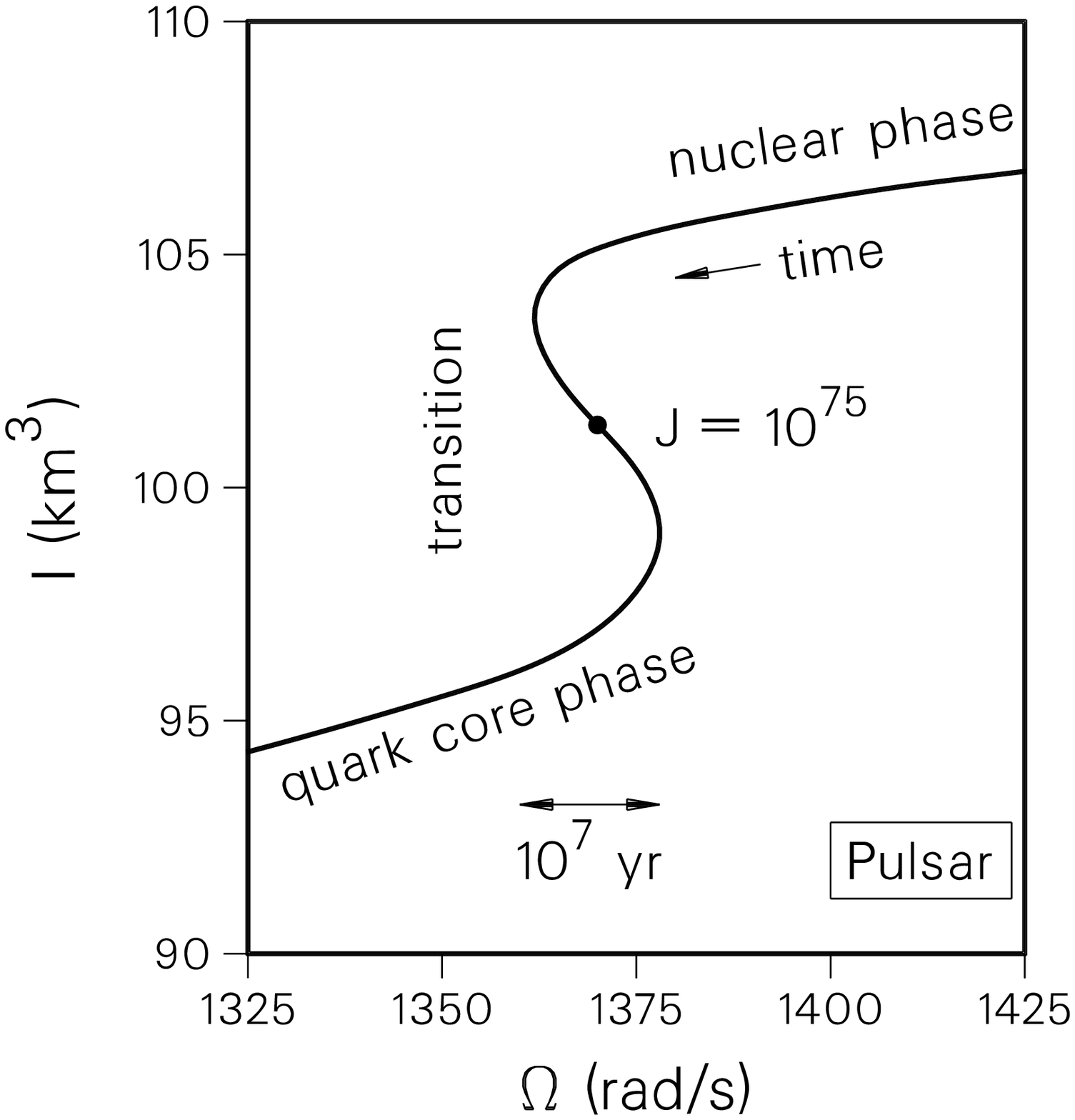}
\includegraphics[width=.5\textwidth]{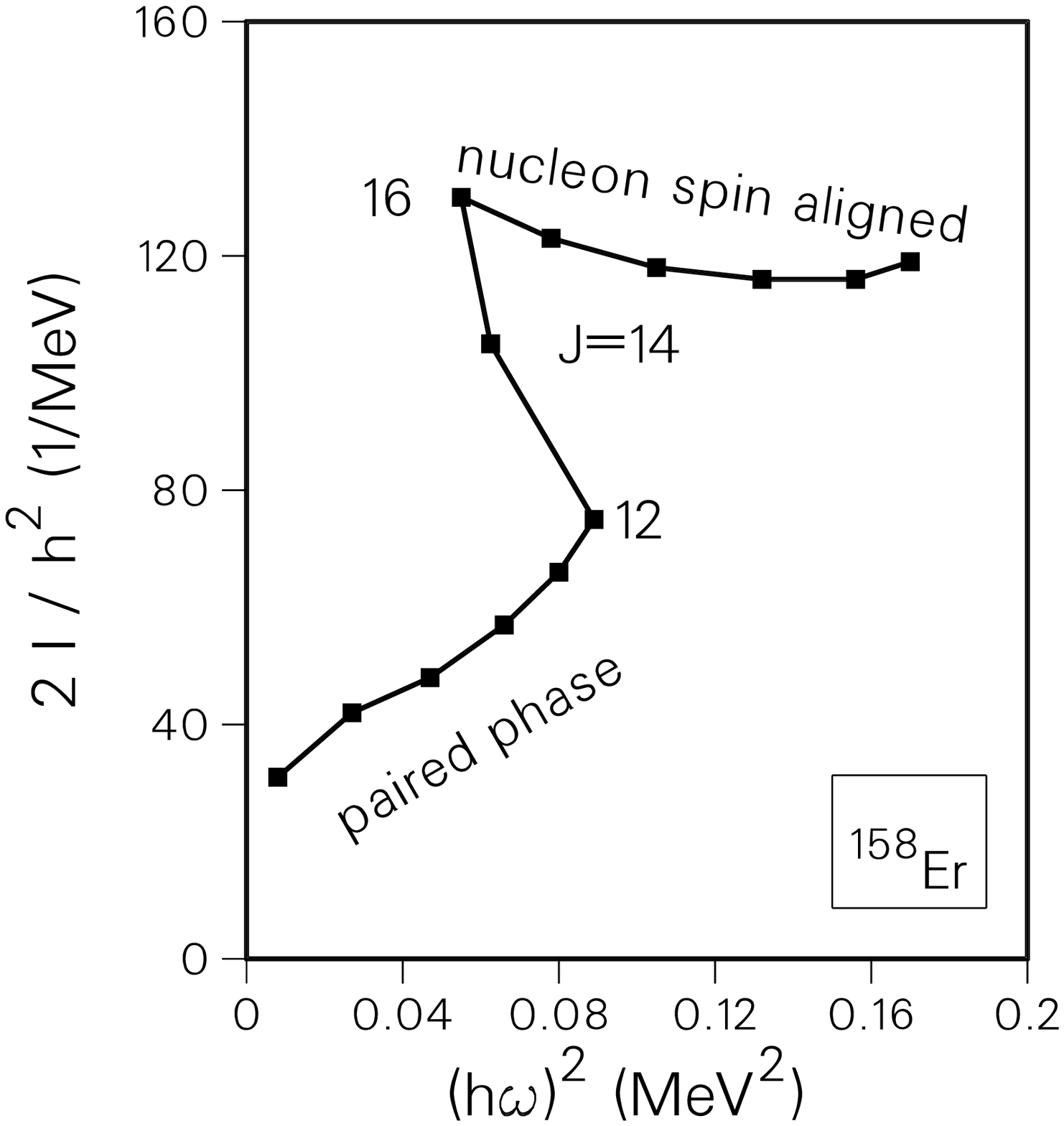}
}}
\begin{flushleft}
\parbox[t]{1\textwidth} { \caption { \label{oif} Moment of 
inertia of a neutron star at angular
velocities for which the central density rises from below to
above critical density for the pure quark matter\index{Quark matter} phase as
the centrifugal force decreases. Time flows from large to small
$I$. The most arresting signal
of the phase change is the spontaneous spin-up that  an
isolated pulsar would undergo during the growth in the
region of pure quark matter.   (Adapted from Ref.\ \cite{glen97:a}.)
}} 
\parbox[t]{1\textwidth} { \caption { \label{nucleusf} Nuclear moment of 
inertia as a function of squared 
frequency for $^{158}$Er, showing backbending in the nuclear case.
Quantization of spin yields the unsmooth curve compared to 
the one in Fig.\
\protect\ref{oif}.
}}
\end{flushleft}
\end{center}
\end{figure}

The behavior of the moment of inertia\index{Neutron star!moment of inertia}
 in the vicinity of the
critical region of growth of the quark 
matter\index{Neutron star!quark matter core} core is shown in Fig.\
\ref{oif}. The critical period of rotation is $P=2\pi/
\Omega\sim 4.6$ ms. Much the same phenomenon 
was observed in rotational nuclei during the
1970s \cite{johnson72:a,stephens72:a} and had been predicted
by Mottelson and Valatin \cite{mottelson60:a} though of course the
phase changes are different
(see Fig.\ \ref{nucleusf}). The mechanism is evidently quite robust,
but it need not occur in every model star. 
In the present instance, the
mass of the neutron star is very close to the limiting mass of the
non-rotating counterpart. This does not necessarily mean that it will
be a rare event. The mass of neutron stars is bounded from below
by the\index{Chandrasekhar limit}
 Chandrasekhar mass limit
which is established by electron pressure
and has a value of $\sim 1.4 M_{\odot}$.
The limiting mass\index{Neutron star!maximum mass}
 of neutron stars may be very little more
because\index{Equation of state!softening} of the softening of the nuclear equation of state by 
hyperonization\index{Hyperonization}
 and quark\index{Quark!deconfinement} deconfinement. This could be the reason\index{Phase transition!deconfinement}
that neutron star masses seem to lie in a very small interval
\cite{books}.

As already pointed out, the progression in time of the growth
of the quark core\index{Neutron star!quark matter core}\index{Neutron star!hybrid} is very slow, being governed by the weak processes 
that cause the loss of angular momentum to radiation. Using the
computed moment of inertia for a star of constant baryon number
as shown in Fig.\ \ref{oif} we can integrate  Eq.\ (\ref{braking2})
to find the epoch of spinup to endure for $2\times 10^7$ y. 
This is a small but significant fraction of the spin-down time
of ms pulsars which is $\sim 10^9$y. So if ms pulsars are near their
limiting mass and are approximately described by our model, about
1 in 50 {\sl isolated} ms pulsars should be spinning up instead of 
spinning down. Presently, about 60 ms pulsars have been identified, 
and about half of them are isolated, the others being in binary
systems. Because the period of ms pulsars can be measured with an 
accuracy that rivals atomic clocks, identification of the
direction of change of period would not take long. 
For example, PSR1937+21 has a period (measured on 29 November 1982
at 1903 UT)
$$P= 1.5578064487275(3) {\rm~ms}\,.$$
Its rate of change of period is a mere $\dot{P}\sim 10^{-19}$.
But because of the high accuracy of the period measurement, it 
would take only
two measurements spaced 
$0.3$  hours apart to detect a unit change in the last significant
figure, and hence to detect in which direction the period is 
changing.
\begin{figure}[tbh]
\vspace{-.24in}
\begin{center}
\leavevmode
\centerline{ \hbox{
\includegraphics[width=.5\textwidth]{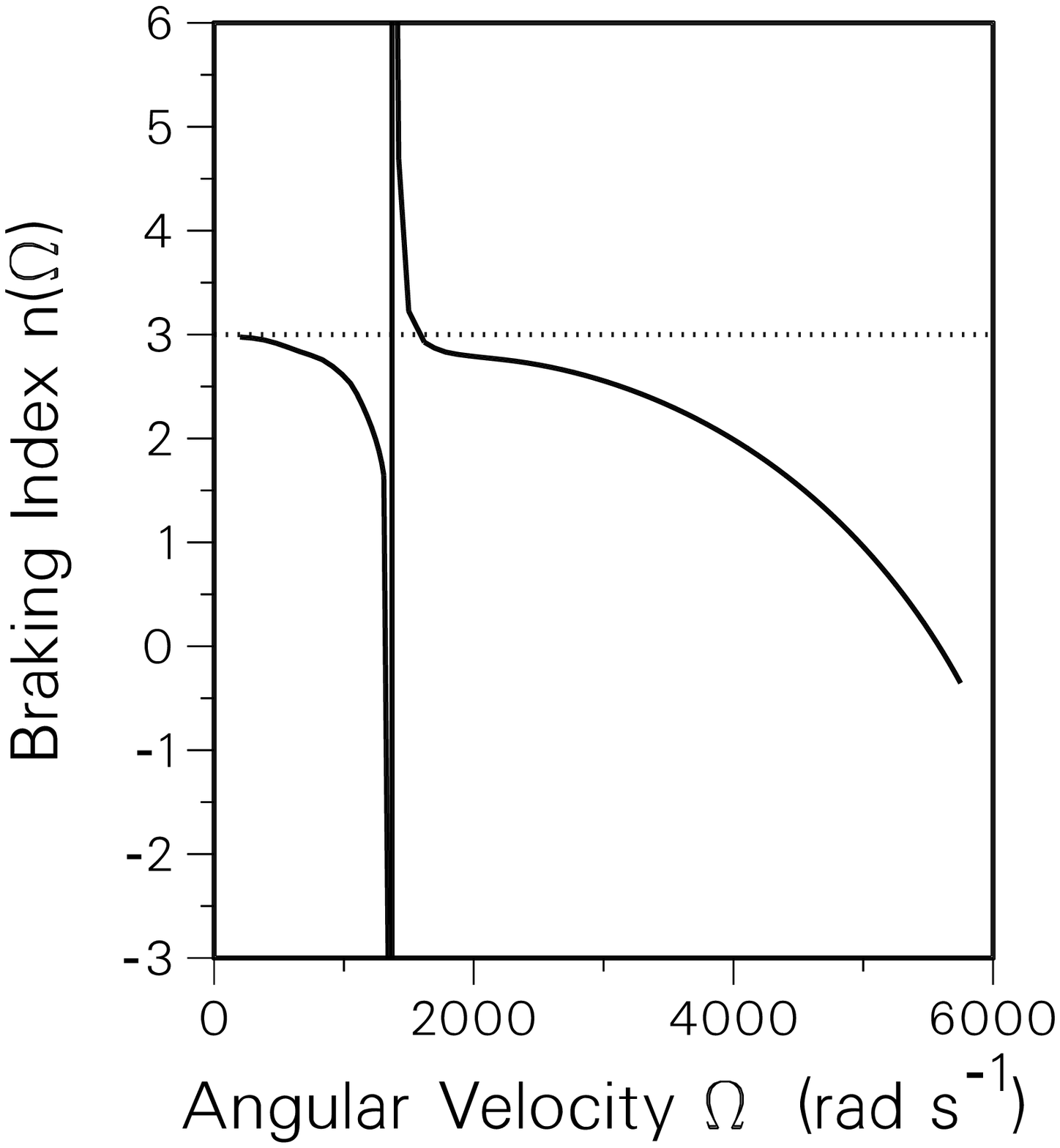}
\includegraphics[width=.5\textwidth]{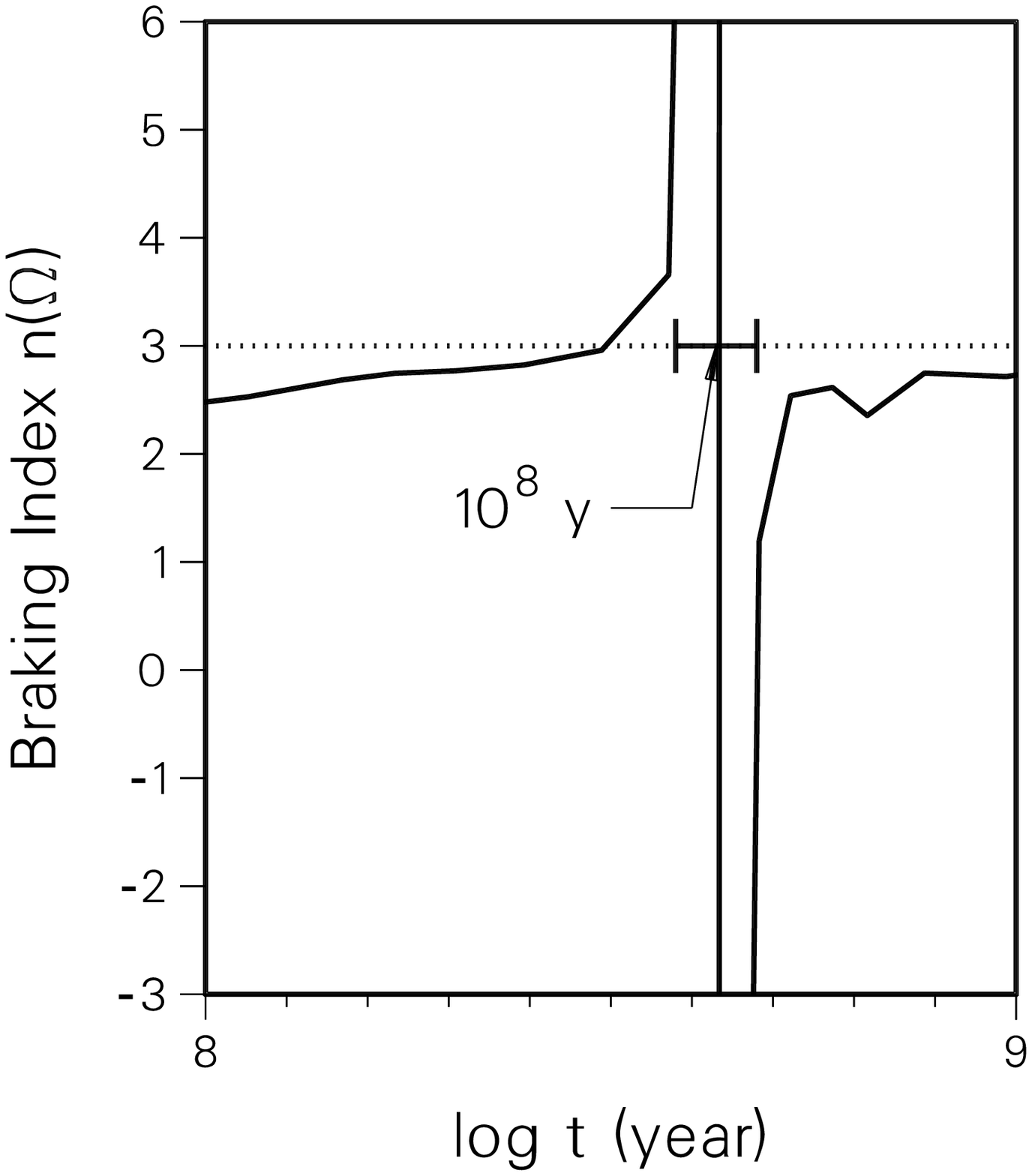}
}}
\begin{flushleft}
\parbox[t]{1\textwidth} { \caption { \label{no1} Braking index
for a pulsar that passes through a change of phase in the central
region as a function of angular velocity\index{Braking index!as 
function of stellar spin}\index{Braking index!during phase transition}.
}} 
\parbox[t]{1\textwidth} { \caption { \label{nt} Braking
index as a function of time over one decade in the critical
region in case of a 
phase transition in the interior of the star. (From Ref.\ \cite{glen97:e}.)
}}
\end{flushleft}
\end{center}
\end{figure}

In principal, the period and first two time derivatives can 
be measured.
The dimensionless ratio formed from them yields the braking index
$n$ of the energy-loss mechanism, corrected by a term that depends
on changes in the moment of inertia. In the case that a phase change 
causes a spinup of the pulsar, as in Fig.\ \ref{oif}, the 
measurable dimensionless quantity 
$n(\Omega)$ has two singularities at the
frequencies at which $dI/d\Omega$ switches between $\pm \infty$.
From Eq.\ (\ref{braking2}) it is clear that $\dot{\Omega}$
will pass through zero and change sign at both turning points. 
Therefore, the measurable braking index Eq.\ (\ref{vindex})
will\index{Braking index!during phase transition} have singularities as shown in Fig.\ \ref{no1}. The braking index
with a nominal value of 3 will make enormous excursions from that
value. The braking index is shown in Fig.\ \ref{nt} over one decade
in time, and has an anomalous value for $10^8$ y. However, 
 the second time
 derivative has never been measured for a ms pulsar
 because
the rate of change of frequency is so slow. So the one, 
and easily observable signal, is the spontaneous spinup of an 
isolated ms pulsar. Such a pulsar has not yet been observed.
But then, only about 30 isolated ms pulsars have so far been 
detected,
whereas, according to our estimate, only 1 in 50 ms pulsars
of mass close to the maximum would presently
be passing through the\index{Neutron star!quark matter core} 
epoch in which a quark matter core grows.

We briefly describe the nuclear\index{Nuclear matter}
 and quark matter\index{Quark matter} 
phases used in this
and the following sections. More details can be found in the Appendix
and in Ref.\ \cite{glen97:a}. The  initial mass of the star 
in our examples is $1.42 M_\odot$.
Briefly,
confined nuclear matter is described by a covariant Lagrangian
describing the interaction of the members\index{Baryon octet}
 of the baryon octet with scalar,
vector and vector-isovector mesons and solved in the meanfield approximation.
Quark matter is described by the MIT bag model.

\section{Accreting X-ray Neutron Stars in Binaries}
\index{Neutron star!x-ray accreter!spin clustering of the population}

Some neutron stars have non-degenerate companions. Such neutron 
stars are 
radio silent because a wind from the hot surface of the 
companion disperses the pulsed radio signal  which a rotating
magnetized neutron star would otherise radiate into space.
Late in the
life of the neutron star, when the slowly evolving
companion begins to overflow 
the Roche lobe, mass transfer
onto the neutron star commences. The drag of the magnetic
dipole torque will be eclipsed by the transfer of mass and angular 
momentum onto the neutron star. It has begun its evolution from an
old\index{Neutron star!evolutionary path}
 slowly rotating neutron star with long period
and high magnetic field, to a ms pulsar 
with\index{Pulsar!recycled}
 low field (sometimes referred to as a ``recycled pulsar'').
 During the long intermediate
stage, when the surface and accretion\index{Neutron star!accretion}
ring\index{Accretion!onto neutron star}\index{Accretion!x-rays from}
 are heated to high temperature, the star radiates x-rays.

Any asymmetry in the mass
accretion pattern will cause a variability
in x-ray emission. At an accretion rate of $10^{-9}\msun /{\rm y}$
it would take 
only $10^8$ y to spin up the neutron star
to a period of 2 ms ($500$ Hz). 
Consequently,
millisecond variability in the x-ray luminosity is expected and observed.

Accreting x-ray neutron stars provide a very interesting contrast to the\index{Neutron star!x-ray accreter} 
spin-down of isolated ms pulsars \cite{glen00:a}. 
Presumably 
they are
the link between the canonical pulsars with mean period of $0.7$ sec
and the ms pulsars \cite{alpar82:a,heuvel91:a,klis98:b,chakrabarty98:a}.
If\index{Phase transition!deconfinement} the critical deconfinement density falls within the range spanned by
canonical pulsars, quark matter will already exist in them but may be ``spun''
out of x-ray stars as their  rotational
frequency increases during accretion.  We can
anticipate that in a certain frequency range, the changing radial extent of the
quark matter phase will actually inhibit changes in frequency because of the
increase in moment of inertia occasioned by the gradual disappearance of the
quark matter phase.  Accreters\index{Neutron star!x-ray accreter}  will tend to spend a greater length of time in
the critical frequencies than otherwise. There will be an anomalous number of
accreters that appear at or near the same frequency. This is what was found
recently in data obtained with the Rossi X-ray Timing Explorer (RXTE).  
 For an extensive review  of the discoveries made
 in the short time since this satellite was launched (1995), see
 Ref.\ \cite{klis00:a}.

The spinup evolution of an accreting\index{Neutron star!x-ray accreter}  neutron 
star is a more complicated problem than\index{Neutron star!accretion}
that of the spindown of an isolated ms pulsar of constant baryon
number. It is complicated by the accretion of matter ($\dot M 
\stackrel{{\textstyle >}}{_\sim}  10^{-10}
M_\odot~{\rm yr}^{-1}$), a changing magnetic field strength (from $B \sim
10^{12} {\rm~to~} \sim 10^8$~G), and the interaction of the field with the
accretion disk.


The change in moment of inertia as a function of rotational frequency
caused by spinup due to accretion\index{Neutron star!x-ray accreter}  is similar to that described in the
previous section, but in reverse \cite{glen97:a}.  However, there are
additional phenomena as just mentioned. The spin-up torque of the
accreting matter causes a change in the star's angular momentum
according to the relation \cite{elsner77:a,ghosh77:a,lipunov92:book}
\begin{eqnarray}
{{d} \over {d t}} J \equiv {{d}\over{dt}} ( I \Omega) = N_A(r_{\rm m}) -
N_M(r_{\rm c}) \, .
\label{eq:dJdt}
\end{eqnarray}
The first term on the right-hand-side is the torque exerted on the
star by a mass element rotating at the base of the accretion disk with
Keplerian velocity $\omega_{\rm K}$.  Denoting this distance with
$r_{\rm m}$, one readily finds that $N_A(r_{\rm m})$ is given by
($G=c=1$)
\begin{eqnarray}
N_A(r_{\rm m}) &=& r_{\rm m}^2 \; {\dot M} \; \omega_{\rm K} \nonumber \\
&=& r_{\rm m}^2 \; {\dot M} \; \left( {{M}\over{r_{\rm
m}^3}} \right)^{1/2} \nonumber  \\
&=& {\dot M} \; \sqrt{M r_{\rm m}} \equiv {\dot M}
\; {\tilde l}(r_{\rm m}) \, ,
\label{eq:torque}
\end{eqnarray}
where $\dot{M}$ stands for the accretion rate, and ${\tilde l}(r_{\rm
m})$ is the specific angular momentum of the accreting matter (angular
momentum added to the star per unit mass of accreted matter).  The
second term on the right-hand-side of Eq.\ (\ref{eq:dJdt}) stands for
the magnetic plus viscous torque term ($\kappa\sim 0.1$),
\begin{equation}
N_M(r_{\rm c}) = \kappa \, \mu^2 \, r_{\rm c}^{-3} \, ,
\label{eq:N}
\end{equation}
with $\mu \equiv R^3 B$ the star's magnetic moment.  Upon substituting
Eqs.\ (\ref{eq:torque}) and (\ref{eq:N}) into (\ref{eq:dJdt}) and
writing the time derivative $d(I \Omega)/dt$ as $d(I \Omega)/dt = (dI/dt)
\Omega + I (d\Omega/dt)$, the time evolution equation for the angular
velocity $\Omega$ of the accreting star can be written as
\begin{equation}
  I(t) {{d\Omega(t)} \over {d t}} = {\dot M} {\tilde l}(t) - \Omega(t)
    {{dI(t)}\over{dt}} - \kappa \, \mu(t)^2 \, r_{\rm c}(t)^{-3} \, .
\label{eq:dOdt.1}
\label{spinevolution} 
\end{equation} 
The quantities $r_{\rm m}$ and $r_{\rm c}$ denote fundamental length
scales of the system. The former,  as mentioned  above,
denotes\index{Accretion!disk}
 the radius of the inner edge of the accretion disk and is
given by $(\xi \sim 1)$
\begin{equation}
  r_{\rm m} = \xi \, r_{\rm A} \, .
\label{eq:r_m}
\end{equation}
The latter stands for the co-rotating radius defined as
\begin{eqnarray}
r_{\rm c} = \left( M \Omega^{-2} \right)^{1/3} \, .
\label{eq:r_c}
\end{eqnarray} Accretion\index{Neutron star!x-ray accreter}  will be inhibited by a centrifugal barrier if the
neutron star's magnetosphere rotates faster than the Kepler frequency
at the magnetosphere. Hence $r_{\rm m} < r_{\rm c}$, otherwise
accretion onto the star will cease. A further fundamental lengthscale
is set by the Alf\'en radius $r_{\rm A}$, which enters in Eq.\
(\ref{eq:r_m}). It is the radius at which the magnetic energy density,
$B^2(r)/8\pi$, equals the total kinetic energy density, $\rho(r)
v^2(r) / 2$, of the accreting matter. For a dipole magnetic field
outside the star of magnitude
\begin{equation}
B(r) = {{\mu}\over{r^3}} \, ,
\label{eq:Br}
\end{equation}
this condition reads
\begin{equation}
{{1}\over{8 \pi}} B^2(r_{\rm A}) = {{1}\over{2}} \rho(r_{\rm A})
v^2(r_{\rm A}) \, ,
\label{eq:alfen.1}
\end{equation}
where $v(r_{\rm A})$ is of the order of the Keplerian velocity, or,
which is similar, the escape velocity at distance $r_{\rm A}$ from the
neutron star,
\begin{equation}
v(r_{\rm A}) = \left( {{2 M} \over {r_{\rm A}}} \right)^{1/2}  \, .
\label{eq:vff}
\end{equation}
The density $\rho(r_{\rm A})$ in Eq.\ (\ref{eq:alfen.1}) can be replaced by
\begin{equation}
 \rho(r_{\rm A}) = {{\dot M} \over{4 \pi r_{\rm A}^2 v^2(r_{\rm A}) }} \, ,
\label{eq:rho}
\end{equation}
which follows from the equation of continuity. With the aid of Eqs.\
(\ref{eq:Br}), (\ref{eq:vff}) and (\ref{eq:rho}), one obtains from
(\ref{eq:alfen.1}) for the Alf\'en radius the following expression:
\begin{equation}
r_{\rm A} = \left( { {\mu^4} \over {2 M \dot{M}^2} } \right)^{1/7} \, .
\label{eq:r_A}
\end{equation} 
It is instructive to write this equation as 
\begin{equation}
r_{\rm A} = 7 \times 10^3 \; {\dot{M}}^{-2/7}_{-10} \; \mu_{30}^{4/7}
\left( {{M}\over{M_\odot}} \right)^{-1/7}~ {\rm km} \, ,
\label{eq:rAkm}
\end{equation}
where $\dot{M}_{-10} \equiv {\dot M}/(10^{-10}~M_\odot~{\rm yr}^{-1})$
and $\mu_{30} \equiv \mu /(10^{30}~ {\rm G}~ {\rm cm}^3)$. It shows
that canonical accreters ($\dot{M}_{-10}=1$) with strong magnetic
fields of $B \sim 10^{12}$~G have Alf\'en radii, and thus accretion
disks, that are thousands of kilometers away from their surfaces.
This is dramatically different for 
accreters\index{Neutron star!x-ray accreter}  whose magnetic fields
have weakened over time to values of $\sim 10^8$~G, for instance.
In this case, the  Alf\'en radius has shrunk to $\sim 40$~km, which is
just a few times the stellar radius.

We assume that the magnetic field\index{Neutron star!magnetic field} evolves according to
\begin{equation}
B(t) = B(\infty) + [B(0) - B(\infty)] e^{-t/t_{\rm d}}
\label{bevolution}
\end{equation} 
with $t=0$ at the start of accretion, and
where $B(0)=10^{12}$ G,~$B(\infty)= 10^8$ G, and $t_{\rm d} = 10^6$ yr.  
Such  a decay to an asymptotic value seems to be a feature of some treatments
of the magnetic field evolution of accreting neutron stars \cite{konar}.
Moreover, it expresses the fact that canonical neutron stars have
high magnetic fields and ms pulsars have 
low fields (see Fig.\ \ref{pulsars123}). Beyond that, as just mentioned,
the\index{Neutron star!accretion}
 condition that accretion can occur demands that
 $r_{\rm m} < r_{\rm c}$ which inequality places an upper limit on the
 magnitude of the magnetic field
of  ms neutron star accreters of $2{\rm~to~}6 \times 10^8$ G
\cite{klis98:b}.

Frequently,
it has been assumed that the moment of inertia 
in Eq.\ (\ref{spinevolution}) does not respond to
changes in the centrifugal force, and in that case, the above 
equation yields a
well-known estimate of the period to which a star can be spun up
\cite{heuvel91:a}.  The approximation is true for slow rotation. However, the
response of the star to rotation becomes increasingly
important as the star is spun up.
Not only do changes in the distribution of matter occur but
internal changes in composition occur also because of changes induced in the
central density by centrifugal dilution \cite{glen97:a}; both changes effect
the moment of inertia and hence the response of the star to accretion. 

\begin{figure}[htb]
\begin{center}
\includegraphics[width=.5\textwidth]{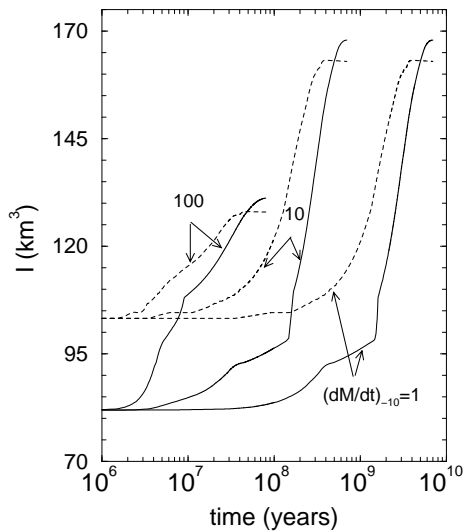}
{ \caption { Moment of inertia of neutron stars as a function of time
with  (solid curves) and without
  (dashed curves) quark matter core
assuming $0.4 M_\odot$ is accreted. Results for 
three {\sl average} accretion rates
are illustrated.
\label{fig:It} 
}}
\end{center}
\end{figure}

The moment of inertia of ms
pulsars or of neutron star accreters\index{Neutron star!moment of inertia} has to be computed in GR
without making the usual assumption of
slow rotation \cite{hartle67:a,hartle68:a}. We use a  previously
obtained  expression for the moment of inertia of a rotating star, good to 
second order in $\Omega$
\cite{glen92:b}.  
The expression is too cumbersome to reproduce
here.  
Stars that are
spun up to high  frequencies close to the breakup
limit (Kepler frequency)
undergo dramatic interior changes;  the
central density may change by a factor of four or so over that of a slowly
rotating star if a phase change occurs
during spin-up (cf. Fig.\ \ref{prof_k300b180})
\cite{glen97:e,weber99:a}.

Figure \ref{fig:It} shows how the moment of inertia changes for neutron stars
in binary systems that are spun up by mass accretion according to Eq.\ 
(\ref{eq:dOdt.1}) until  $0.4 M_{\odot}$
 has been accreted. The neutron star models are fully
described in 
Ref.\ \cite{glen97:a} and references therein and  briefly
in the Appendix
of this article.
The initial mass of the star in our examples is $1.42 M_\odot$.
In one case, it is assumed that a phase transition between
quark matter and confined hadronic matter occurs, and
in the other that it does not. This accounts for the different initial moments
of inertia, and also, as we see, the response to spinup.
Three accretion rates\index{Accretion!rate} are assumed, which range from
$\dot{M}_{-10}=1$ to 100
(where $ \dot{M}_{-10}$ is measured in units
of $10^{-10} M_{\odot}$ per year).  
These rates are in accord with observations\index{Binaries!x-ray} 
 made on
low-mass X-ray binaries (LMXBs) observed with the Rossi X-ray Timing Explorer
\cite{klis00:a}. The observed objects, which are divided into Z sources and
A(toll) sources, appear to accrete at rates of $\dot{M}_{-10} \sim 200$ and\index{Accretion!rate}
$\dot{M}_{-10} \sim 2$, respectively.
Although in a given binary, $\dot{M}$ varies on a timescale of days, 
we take it to be the constant average rate in our calculations.

\begin{figure}[htb]
\begin{center}
\includegraphics[width=.5\textwidth]{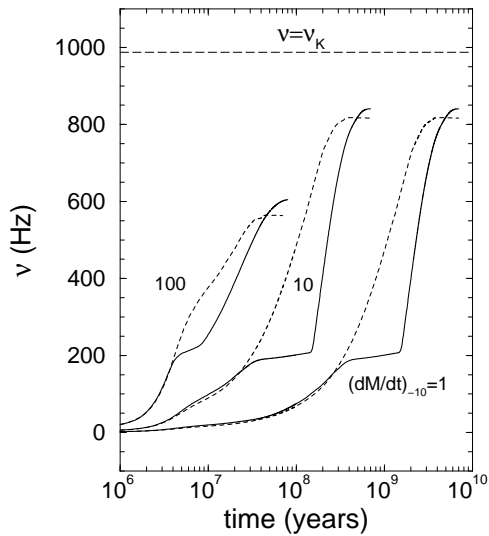}
{ \caption {Evolution of spin frequencies ($\nu \equiv \Omega / 2\pi$)
of accreting neutron stars with (solid curves) and without (dashed
curves) quark deconfinement\index{Phase transition!deconfinement} if $0.4 M_\odot$ is accreted.
(If the mass of the donor star is less, then so is the maximum
attainable frequency.)
The spin plateau around $200$~Hz
signals the ongoing process of quark confinement in the stellar
centers.  Note that an equilibrium spin is eventually reached which is less
than the Kepler frequency.
\label{fig:nue}
}}
\end{center}
\end{figure}

Figure \ref{fig:nue} shows the spin evolution of accreting neutron stars as\index{Neutron star!spin!evolution}
determined by the changing moment of inertia and the spin 
evolution equation,
Eq.\
(\ref{spinevolution}).  Neutron stars 
{\sl without} quark matter\protect\index{Neutron star!quark matter core} in their centers are
spun up along the dashed lines to equilibrium frequencies between about 600~Hz
and 850~Hz, depending on accretion rate and magnetic field. The $dI/dt$ term
for these sequences manifests itself only insofar as it limits the equilibrium
periods to values smaller than the Kepler frequency,
$\nu_{\rm K}$.  
In both Figs.\ \ref{fig:It} and  \ref{fig:nue} we assume that $0.4 M_{\odot}$
is accreted. Otherwise, the maximum frequency attained is less, the less
matter is accreted.

The spin-up scenario is dramatically different for neutron stars in 
which a first
order phase transition occurs.
In this case, as known from Fig.\ \ref{fig:It}, the
temporal conversion of quark matter into its mixed phase of quarks and confined
hadrons is accompanied by a pronounced increase of the stellar moment of
inertia. This increase contributes so significantly to the torque term
$N(r_{\rm c})$ in Eq.\ (\ref{eq:dOdt.1}) that the spinup rate $d\Omega/dt$ is
driven to  a plateau around those frequencies at which the pure quark matter
core in the center of the neutron star gives way to the mixed phase of confined
hadronic matter and quark matter.  The star resumes ordinary spin-up 
when this
transition is completed.  The epoch during which the spin rates reach a plateau
are determined by attributes like the accretion rate, magnetic field, and its
assumed decay time. The epoch\index{Accretion!epoch}
 lasts between $\sim 10^7{\rm~and~}10^9$ yr
depending on the accretion rate at the values taken for the other factors.

\begin{figure}[htb]
\begin{center}
\includegraphics[width=.55\textwidth]{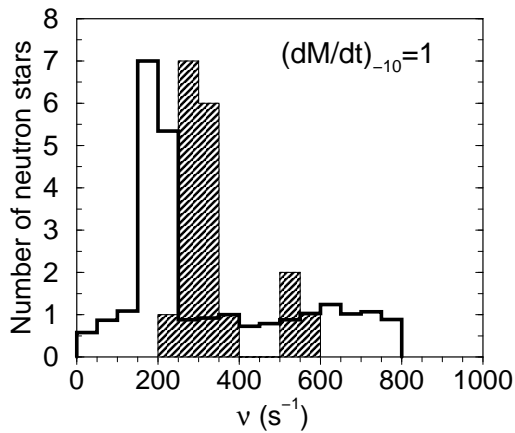}
{ \caption { Frequency distribution of X-ray neutron stars. Calculated
distribution (open histogram) for the underlying population is
normalized to the
number of observed objects (18)
at the peak. (The normalization causes a fractional number
to appear in many bins of the calculated distribution.)
 Data on  neutron stars in
 LMXBs (shaded histogram) is from Ref.\  \protect\cite{klis00:a}. See
 text and especially the reference for caveats to the interpretation.
 The spike in the calculated distribution
 corresponds to the spinout of the quark matter phase and the
 corresponding growth of the moment of inertia as compressible
 quark matter is replaced by relatively incompressible nuclear
 matter.  Otherwise the spike
 would be absent. \label{fig:bin}
 }}
 \end{center}
 \end{figure}

We can translate the information in Fig.\ \ref{fig:nue} into a frequency
distribution of X-ray stars 
by\index{Neutron star!spin!distribution of accreters}
assuming that neutron stars begin their accretion
evolution at the average rate of one per million years. A different rate will
only shift some neutron stars from one bin to an adjacent one, but
will not change the basic form of the distribution.  
The donor\index{Binaries!donor mass}  masses in the binaries\index{Binaries!x-ray}  are believed to range between
 $0.1 {\rm~and~} 0.4 M_{\odot}$. For lack of more precise knowledge,
 we assume a uniform distribution of donor masses (or mass accreted)
 in this range and repeat
 the calculation shown   in Fig.\ \ref{fig:nue} at intervals of
  $0.1 M_{\odot}$.

The result for the computed distribution of rotational frequency
of x-ray neutron stars
is
shown in Fig.\ \ref{fig:bin}; it  is striking. Spinout of the
quark matter core\index{Neutron star!quark matter core} 
as the neutron star spins up is signalled by a spike
in the  distribution which would be absent
if there were no phase transition in our model of the neutron
star.  We stress that
what we plot is our prediction of the {\sl relative}
frequency distribution
of the  {\sl underlying} population of x-ray neutron stars---but
the weight given to the spike
as compared to the high frequency
tail depends sensitively on the weight with which
the donor masses are assigned and the initial mass distribution
of neutron stars in LMXBs. As already mentioned, we give equal weight
to donor masses between $0.1 {\rm~and~} 0.4 M_{\odot}$ and the initial
mass of the neutron star is $1.42 \msun$.
The objects above 400 Hz in Fig.\ \ref{fig:bin} are actually unstable
and will collapse to black holes. Donors of all masses in the range
just mentioned contribute to neutron stars of spin up to 400 Hz.
Neutron stars of lower initial mass than our $1.42 \msun$
 with donors of mass at the
higher end of their range will produce spins above 400 Hz. In other words,
the relative population in the peak as compared to the background
will be sensitive to the unknown factors (1) accretion rate (2)
initial mass distribution of neutron stars in LMXBs (3) mass distribution of donor stars. However, the position of the peak
in the spin distribution of x-ray neutron stars is a property
of nuclear matter and independent of the above unknowns.

The calculated concentration  in frequency of x-ray neutron stars  is
centered around 200 Hz;  this is about 100 Hz lower than the observed spinup
anomaly (see discussion below). 
This discrepancy should not be surprising in view of our total
ignorance\footnote{There are upcoming radioactive beam experiments
from which it is hoped to gain information on the equation of state~ of asymmetric\index{Equation of state} 
nuclear matter \protect\cite{tanihata,hansen,bao}.}
of the equation of state~ above saturation density of nuclear matter and the
necessarily crude representation of hadronic matter in the two phases
in the absence of relevant solutions to the fundamental QCD theory of
strong interactions. We represent the confined phase
by relativistic nuclear field theory and  the deconfined phase by
the MIT bag 
model. However crude these or any other models of hadronic matter may be,
the physics underlying the effect of a phase transition on
spin rate is robust, although not inevitable.
We have cited the example of an analogous
phenomenon  found in rotating nuclei in the previous section.

\begin{figure}[t]
\begin{center}
\includegraphics[width=.55\textwidth]{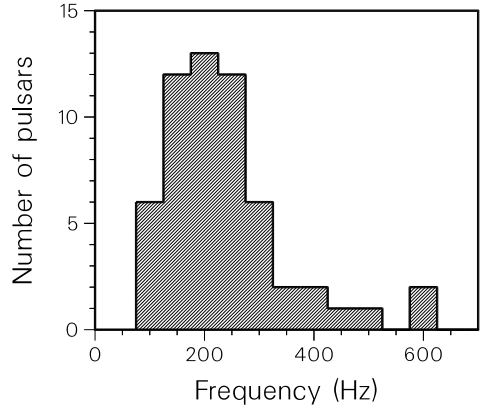}
\end{center}
\caption[]{Data on frequency distribution of millisecond pulsars
($1 \leq P <  10$ ms). Frequency bins are 50 Hz wide.
}
\label{fig:princeton}
\end{figure}

The data that we have plotted in Fig.\ \ref{fig:bin}
is gathered from Tables 2, 3, and 4 of the review article of
van der Klis concerning discoveries made with 
the  Rossi X-ray Timing Explorer, launched near the end of 1995
\cite{klis00:a}. The interpretation of millisecond oscillations
in the x-ray emission, either that  found in  bursts or of the difference\index{X-ray!bursts}
between two observed quasi-periodic oscillations
(QPOs)\index{QPO}\index{X-ray!quasi periodic oscillations} in x-ray brightness,
is  ambiguous in some cases.  In particular, the highest frequency near
600 Hz in the ``observed'' data displayed in our
figure, may actually be twice the rotational
frequency of the star \cite{klis00:a}. 
The millisecond variability in x-ray phenomena associated with accretion 
onto neutron stars that has been observed since the launch of the satellite
was anticipated several decades ago. However, the consistency
of the phenomena from one binary to another raises questions about 
interpretation. In this sense, the field is quite young.
We refer to the above cited review article for details and references to
the extensive literature.

Nevertheless, the basic feature will probably survive---a clustering
 in rotational frequencies\index{Neutron star!x-ray accreter!spin clustering of the population} 
of x-ray neutron stars  and a higher frequency
tail.
Certainly there are high frequency {\sl pulsars}.
  A histogram
of {\sl ms pulsar} frequencies shows
 a broad distribution around  200 Hz, and a  tail extending to $\sim 600$ Hz
 as shown in Fig.\ \ref{fig:princeton}.
So both the (sparse) data on X-ray
objects and on ms pulsars seem to  agree on a peak 
in the number of stars at moderately high rotational
frequency
 and on  an attenuation at
 high
frequency. 
For ms pulsars, however, the attenuation at high frequency may be  
partly a selection\index{Interstellar medium!dispersion of radio signal}
effect due to interstellar disperion of the radio signal.

There have been other  suggestions 
as to the\index{Neutron star!x-ray accreter!spin clustering of the population}
cause of the spike, several of which
we cite (c.f. \cite{bildsten98:a,anderson2000:a,levin99:a}). 
These works are concerned with the balance of the spinup
torque by a gravitational radiation torque.
Our proposal has several merits: (1) the mechanism involving
a change in moment of inertia triggered by a phase transition in 
the stellar core
due to changing density profile in the star as its spin changes
is robust---it is known to occur in rotating nuclei;
(2) The phase transition should occur in reverse in
isolated ms pulsars and in neutron star accretors and at about
the same frequency for similar mass stars;
(3) The phase transition causes accretion induced spinup to stall
for a long epoch, but to resume after the quark core has
been expelled, thus accounting for high spin objects like
very fast ms pulsars as well as the clustering in spin of the
popuation of accretors.

\section{Evolution from Canonical to Millisecond Pulsar}

In\index{Pulsar!evolution from canonical to ms pulsar}
 the foregoing we have discussed possible signals of a phase 
transition in isolated ms pulsars and in accreting x-ray neutron 
stars\index{Neutron star!in low-mass x-ray binary}
 in binary orbit with a low-mass companion. Spinup by 
mass accretion\index{Neutron star!accretion}
 is believed to be the pathway from the relatively
slowly\index{Accretion!pathway from canonical to ms pulsar}
 rotating\index{Neutron star!evolutionary path}
 canonical pulsars formed by conservation
of angular momentum in the core collapse of massive stars,
 and the rapidly rotating
millisecond pulsars \cite{alpar82:a,heuvel91:a}. 
In this section we trace some of the possible evolutionary routes
under the various physical conditions under which accretion occurs
 \cite{glen00:b}.

The evolutionary track between canonical and ms pulsars in the
coordinates of magnetic field strength and rotational period
(refer to Fig.\ \ref{pulsars123})
will  depend on the rate of mass accretion, its duration,
the centrifugal
change in the moment of inertia of the star,
the strength of the
 magnetic field\index{Neutron star!magnetic field!evolution}
 and timescale of its decay, and possibly other 
affects.
Many papers have been devoted to the decay
of the magnetic field. It is an extremely complicated
subject, with many physical uncertainties such as the
actual location of the field, whether in the core or crust,
the degree to which the crust is impregnated with impurities, crustal
heating and resultant reduction in conductivity and 
therefore increase in
ohmic  field decay, screening of the magnetic field by accreted material,
and so on.

The field is believed to decay only weakly due to ohmic
resistance in canonical pulsars, but very significantly
if in binary orbit with a low-mass non-degenerate star,
when the companion fills its Roche lobe. This era can last
up to $10^9$ y and cause field decay by several orders of magnitude.
For a review of the literature and several evolutionary 
scenarios, see Refs.\  \cite{lipunov92:book,konar,urpin96,urpin97,urpin98}. 
 
While there
is no consensus concerning the magnetic field decay,
observationally, we know that canonical
pulsars have fields of $\sim 10^{11} {\rm~to~} 10^{13} G$, while
millisecond pulsars have fields that lie in the range
$\sim 10^{8} {\rm~to~} 10^{9} G$. We shall rely on this observational
fact, and assume that the field decays according to 
Eq.\ (\ref{bevolution}).
where  $B(\infty)=10^8$ G,
$B(0)=10^{12}$ G and $t_{\rm d} = 10^5$ 
to $10^7$ yr. Moreover, this is the general form found in some
scenarios  \cite{konar}.  
However, we shall also make a comparison with a
purely exponential decay. 

There are three distinct aspects to developing an evolutionary 
framework.
One has to do with the accretion process itself, which has been developed 
by a number of authors in the framework of classical 
physics(\cite{elsner77:a,ghosh77:a,lipunov92:book})
and which we employed in the previous section. Another has to do with
the field decay, which in a complete theory will be coupled to
the accretion process.
The third
 aspect has to do with\index{Neutron star!structure} the structure of the neutron star and 
 its response to added mass, but most especially to 
its response to changes in rotational frequency due to the changing 
centrifugal forces.

Typically, the moment of inertia has been computed in general 
relativity for a non-rotating star \cite{hartle67:a,hartle68:a}.
It is based on the Oppenheimer-Volkoff metric. However, for the 
purpose of tracing the evolution of an accreting star from 
$\sim 1$ Hz 
to $400-600$  Hz we do not
neglect\index{Neutron star!structure} the response of the shape, structure and composition 
of the star as it is spun up over this vast range of  
frequencies from essentially zero to values that approach
the Kepler frequency. Nor do
we neglect the dragging of local inertial frames.
These features are included in our calculation of the
tracks of neutron stars from canonical objects starting with 
large fields and very small frequencies at the ``death line'' 
to the small fields but rapid rotation of millisecond pulsars.  
However, the expression
of the moment of inertia and a definition of the various factors 
that enter are too long to reproduce here. We refer instead to our 
derivation
given in Refs.\  \cite{glen92:b,blaschke99:a} which is computed to 
second order in the rotational angular momentum and found to be
accurate to $\sim 10$ \% when compared to numerical solutions
for a rotating star
\cite{haensel94:a}. In the present context, numerical solutions
were obtained in Ref.\  \cite{cook}, and semianalytic
approximations were employed in Ref.\ \cite{burderi}.  
\begin{figure}[htb]
\begin{center}
\includegraphics[width=.5\textwidth]{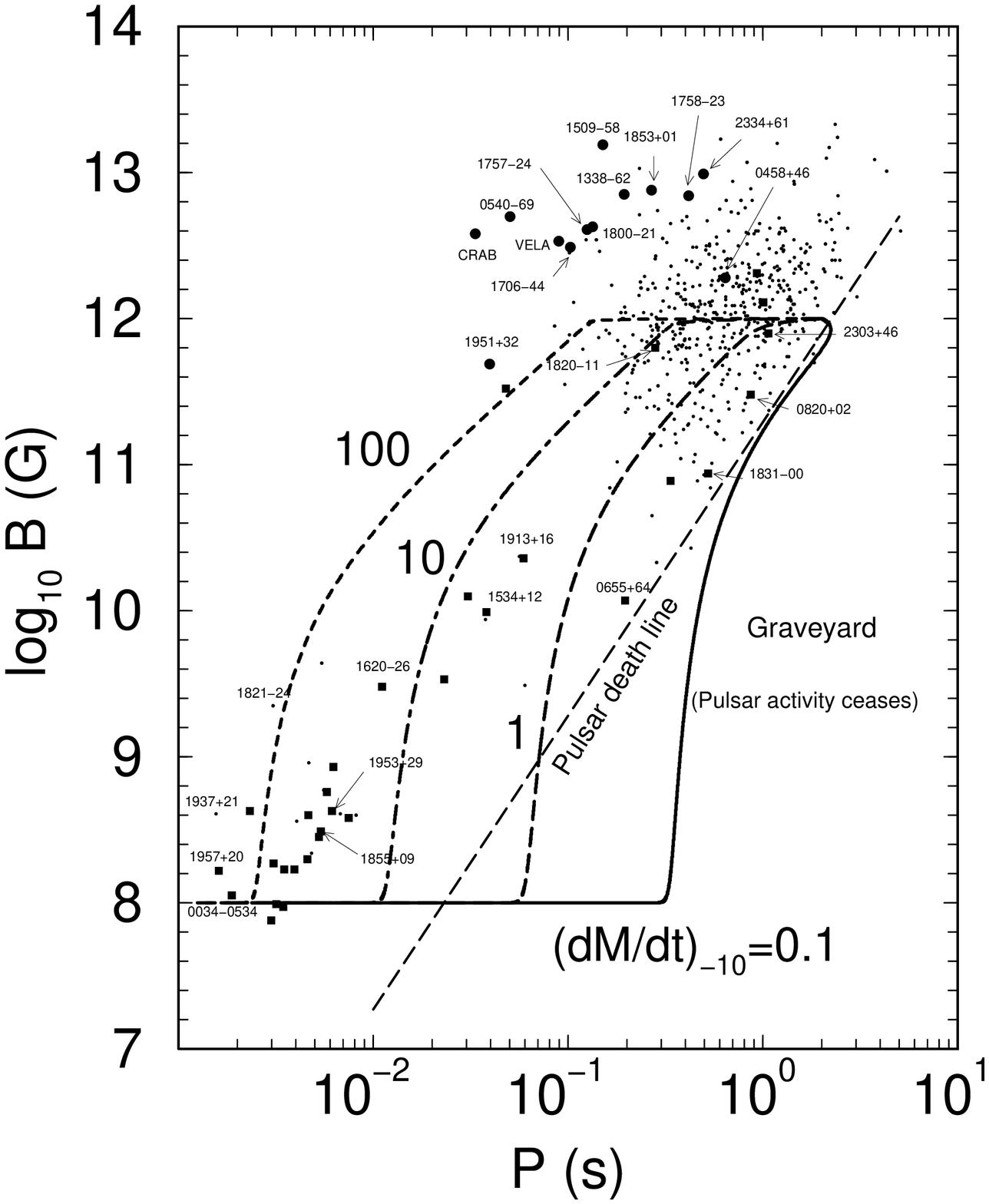}
\end{center}
\caption { \label{bp_4} Evolutionary tracks traced by neutron stars
in the X-ray accretion stage, 
beginning 
on the death line with large $B$ field
and ending as  millisecond stars, for various accretion rates.
Here $t_{{\rm d}}=10^6$yr.
}
\end{figure}

The initial
conditions for the evolution are arbitrary to a high degree.
Canon\-ic\-al pulsars have a broad range of magnetic field strengths.
The period of the pulsar
at the time that the companion overflows\index{Roche lobe}
 its Roche lobe and accretion\index{Neutron star!accretion}
commences is also arbitrary. Any observed sample of x-ray accreters
presumably spans a range in these variables. For concreteness,
we assume that the pulsar has a field of $10^{12}$ G and that the
period of the pulsar is 1 ms when accretion begins. The donor mass
in the\index{Accretion!donor mass}
 low-mass binaries are in the range $0.1{~\rm to~}0.4 M_{\odot}$.
Our sample calculations are for accretion of up to $0.4M_{\odot}$.

\begin{figure}[htb] \begin{center}
\includegraphics[width=.5\textwidth]{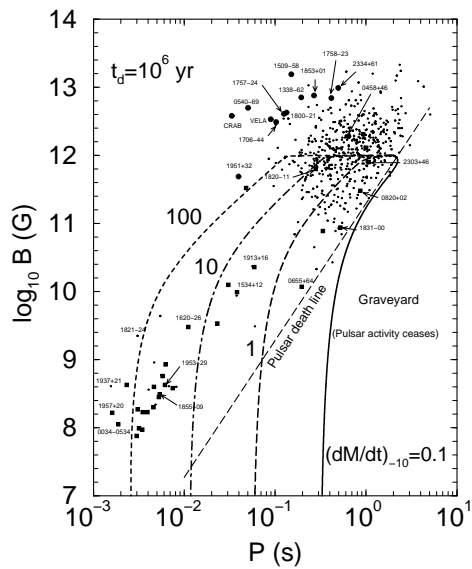}
\end{center} { \caption { \label{bp_4e} Evolutionary\index{Neutron star!evolutionary path} tracks traced by 
neutron stars in the X-ray accretion stage, beginning on the death 
line with purely exponential decay of the $B$ field. As in Fig.\ 
\protect\ref{bp_4}, $t_{{\rm d}}=10^6$yr. }} \end{figure}

We find essentially
 a continuum of evolutionary tracks in the $B-P$ plane
according to the rate at which matter is accreted from the companion, and the 
rate at which the magnetic field decays.  The evolutionary tracks essentially 
fill all the space in the B-P plane, starting at our assumed initial 
condition of an
old canonical pulsar with field of $10^{12}$ G, and extending 
downward in field strength, broadening to fill the 
space on both sides of the deathline, and extending to the 
small periods of  millisecond pulsars.     
All are potential tracks of 
some particular binary pair, since accretion rates vary by several 
orders of magnitude and 
presumably so do decay rates of the magnetic field.    
As a first orientation as to our results and how they relate to  known 
pulsars as regards their magnetic field strength and their rotational 
period, we show the evolutionary tracks for four different accretion 
rates given in units of $\dot{M}=10^{-10}$ solar masses per year in  
Fig.\ \ref{bp_4}. The decay rate of the magnetic field is taken to have 
the value $ t_{{\rm d}}=10^6$ yr in each case. The x-ray neutron star 
gains angular momentum and its period decreases, and over a longer  
timescale, the magnetic field decays. One can see already that a wide  
swathe of $B$ and $P$ is traced out.

In the above example, the field was 
assumed to decay to a finite asymptotic value of $10^8$ Gauss.  A very 
different assumption, namely that the field decays eventually to zero, 
$B(t)=B(0)e^{-t/t_{{\rm d}}}$, modifies only the results below the asymptotic 
value, as is seen by comparing Fig.\ \ref{bp_4} and \ref{bp_4e}.  However, 
the conclusion concerning the origin of millisecond pulsars is  quite 
different. For purely exponential decay, one would conclude that 
high frequency pulsars are created only in high accretion rate binaries.

In the remainder 
of the paper, we assume the field decays to an asymptotic  value, 
since from the above comparison we see how exponential decay would 
modify the picture. 

\begin{figure}[tbh]
\begin{center}
\leavevmode
\centerline{ \hbox{
\includegraphics[width=.44\textwidth]{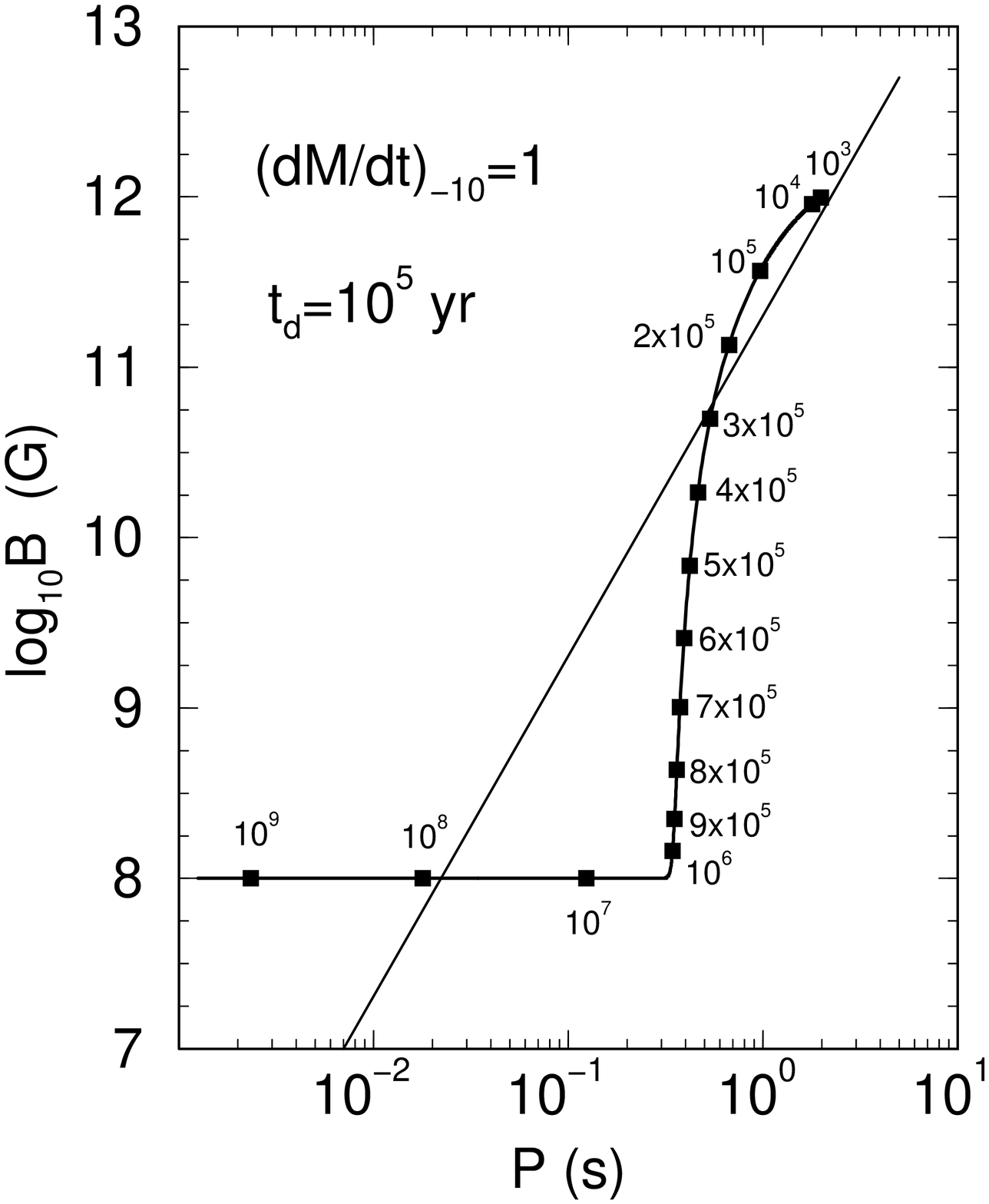}
\hspace{.4in}
\includegraphics[width=.44\textwidth]{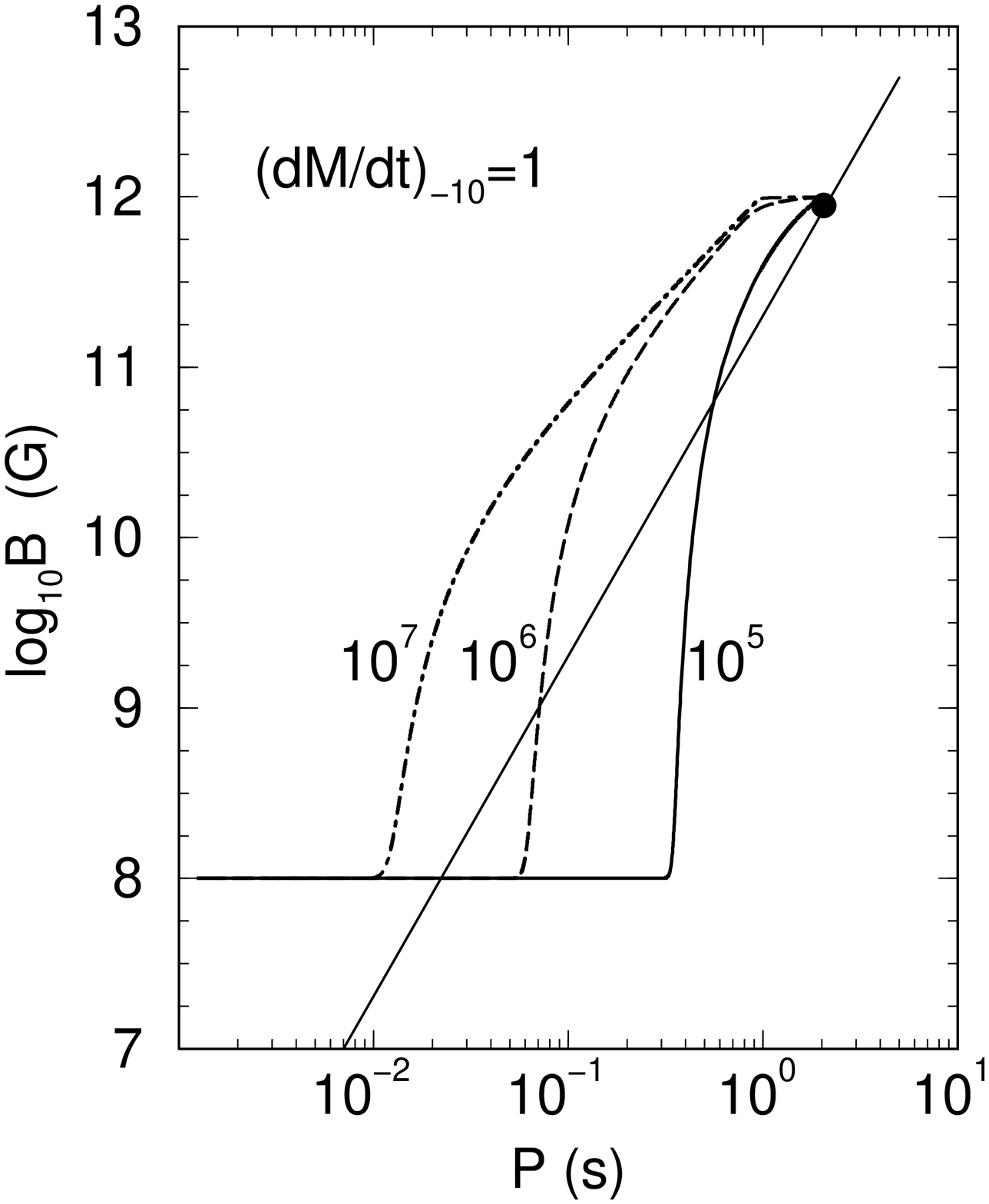}
}}
\begin{flushleft}
\parbox[t]{1\textwidth} { \caption { \label{time_tag} Time tags 
expressed in years are shown for one of the
evolutionary tracks corresponding to a decay constant for the 
magnetic field of $t_d=10^5$ y. 
}} 
\parbox[t]{1\textwidth} { \caption { \label{bp_a} Evolutionary 
tracks for a neutron star
starting at the death line and evolving by accretion to lower
field and high frequency for an accretion rate 1 in units of 
$10^{-10}$ solar masses per year, and for three values of the
magnetic field decay rate $t_{{\rm d}}$ as marked.
 }}
\end{flushleft}
\end{center}
\end{figure}

\begin{figure}[tbh]
\begin{center}
\leavevmode
\centerline{ \hbox{
\includegraphics[width=.44\textwidth]{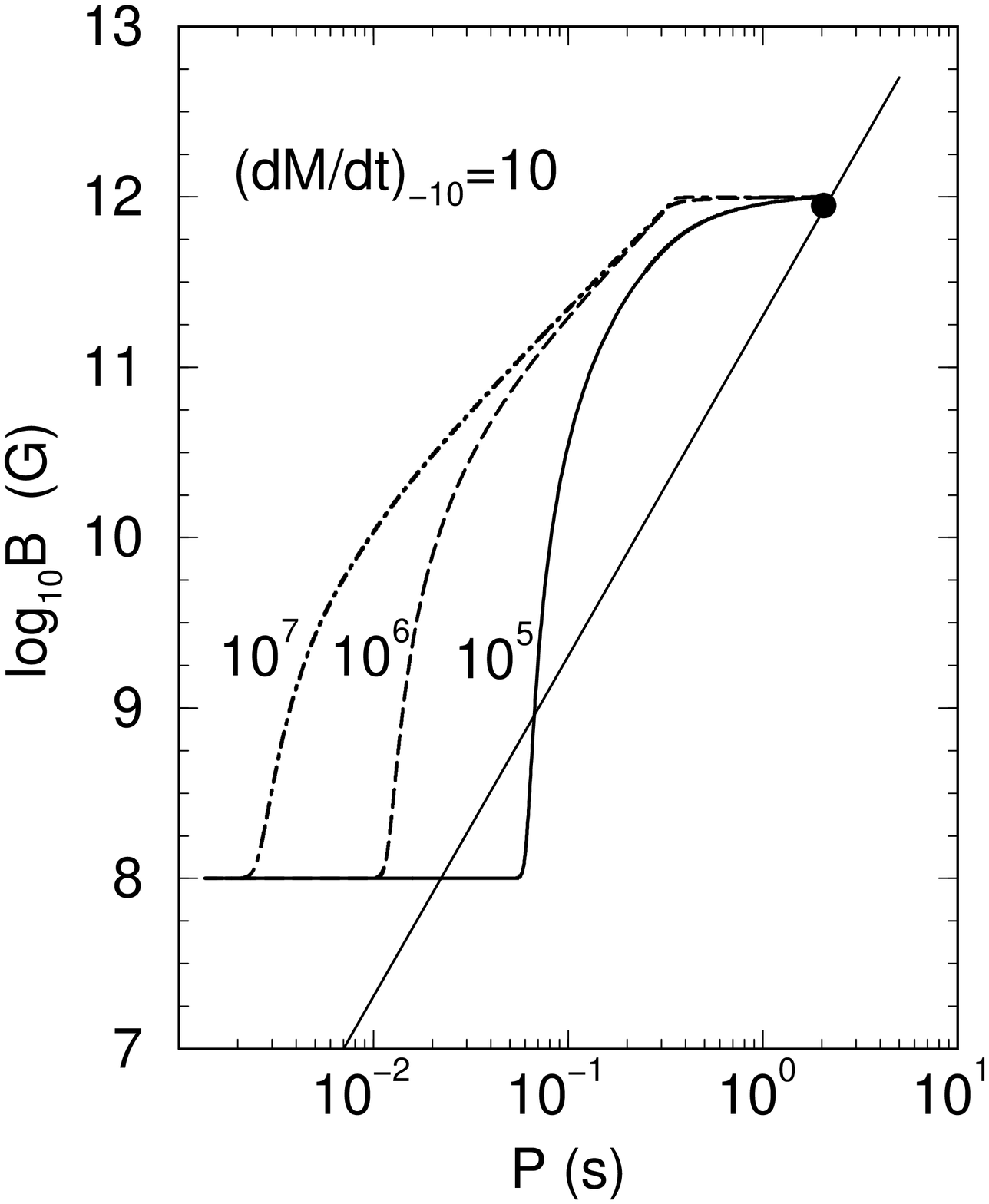}
\hspace{.4in}
\includegraphics[width=.44\textwidth]{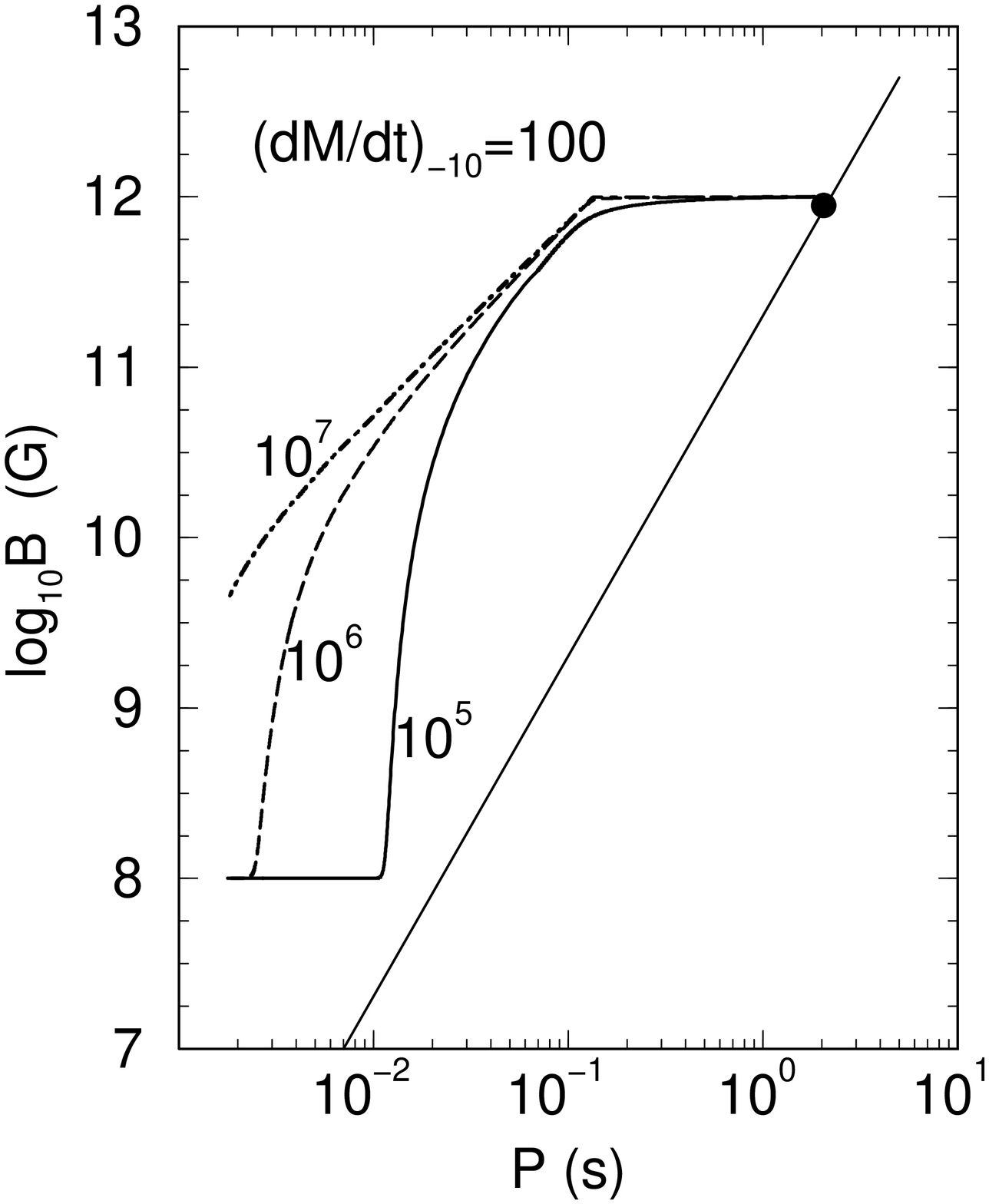}
}}
\begin{flushleft}
\parbox[t]{1\textwidth} { \caption { \label{bp_b} Evolutionary
tracks similar to Fig.\ \protect\ref{bp_a}
but with a different accretion rate $(dM/dt)_{-10}=10$.  
}} 
\parbox[t]{1\textwidth} { \caption { \label{bp_c}  Evolutionary
tracks similar to Fig.\ \protect\ref{bp_a}
but with a different accretion rate $(dM/dt)_{-10}=100$. 
}} 
\end{flushleft}
\end{center}
\end{figure}

We show time tags on a sample track in 
Fig.\ \ref{time_tag} which provides some sense of time lapse. 
The first part of a track is traversed in short time, but the remainder 
ever more slowly. This shows up also in $dP/dt$ as a function of time.   
For each of three accretion\index{Neutron star!accretion}
 rates we show the dependence on three  field 
decay constants in Figs.\ \ref{bp_a}, \ref{bp_b} and \ref{bp_c}. Depending 
on decay rate of the field and accretion rate, an X-ray neutron star may 
spend some time on either side of the death line, but if it accreted long 
enough, always ends up as a candidate for a millisecond pulsar {\sl if} the 
magnetic filed decays to an asymptotic value such as was assumed. However, 
if the field decays exponentially to zero, only high accretion rates would 
lead to  millisecond pulsars. Of course, if accretion turns off at some 
time,  the evolution is arrested.  

In summary, we have computed the evolutionary tracks  in the $B-P$ 
plane due to mass accretion onto  neutron stars beginning at the death line 
with a typical field  strength of $10^{12}$ Gauss, to shorter periods and low 
fields.  According to the assumed accretion rate and decay constant for the 
magnetic field, the tracks indicate that the individual binaries with 
characteristics ranging from Z to Atoll sources will evolve along paths that 
cover a broad swathe in the $B-P$ plane. These  include tracks of X-ray stars 
corresponding to low accretion rates  that follow a path beyond the death 
line in\index{Pulsar!graveyard} the so-called ``graveyard'.  We have assumed two  particular forms 
for the law of decay of the magnetic field. (1) The field  approaches an 
asymptotic value of $10^8$ Gauss such as is typical of millisecond pulsars. 
This assumption leads to a  particular form for the termination of 
evolutionary tracks\index{Neutron star!evolutionary path}.
 All accreters, no matter what the accretion rate,  
will end with millisecond periods, unless accretion ceases beforehand.  
(2) If instead, we had assumed a purely exponential decay,  the tracks would 
not tend to an asymptotic value,  but would continue to decrease in the 
strength of $B$. The tracks would  still cover a broad swathe in the $B-P$ 
plane.  But one would conclude that  only the higher accretion rate binaries,\index{Binaries!x-ray}  
particularly the Z-sources, could produce millisecond period neutron stars. 
If accretion continues for too long a time, the neutron star will be carried 
to very low fields and across the death line, or an overcritical mass 
will have been accreted, leading instead to a black hole.   

\appendix

\section{Appendices}

\subsection{First Order Phase Transition in Neutron Stars}

We briefly recall some of
the\index{Phase transition!first order} 
main characteristics of a first order
phase transition in any substance having more than one conserved
quantity, or\index{Conservation!of baryon number}\index{Conservation!of charge} as we will call it in the context of physics---conserved
charge---such as baryon number or electric charge
\cite{glen91:phase}. These are the\index{Neutron star!conserved charges}
two conserved quantities relevant to neutron stars. They have
zero net electric  charge and are made from baryons. 

What makes a substance having more than one conserved charge
different from a substance having only one, is that 
charges in the first case can be exchanged between two phases
of the substance in equilibrium so as to minimize the energy. 
And the concentration of the charges can readjust at each proportion
of the phases to minimize the energy. There is no such degree of
freedom\index{Phase transition!degrees of freedom}
 in a single-component substance, and there are $n-1$ degrees
of freedom in a substance having $n$ independent components or
conserved charges. 

In a single-component substance, the pressure at constant temperature
remains unaltered for all proportions of the two phases as 
the substance is compressed. Only the proportion of the phases
changes.
This is
not so for a substance of more than one conserved charge. The 
degree(s)
of freedom to 
readjust  the concentrations  of the charges
at each proportion of the phases
causes a change in all internal
properties of the two phases as their
proportion changes including their  common pressure under conditions
of constant temperature. 

The fact that the pressure varies as the proportion of phases in 
equilibrium in neutron star matter has an immediate consequence. 
The mixed phase of two phases in equilibrium will span a finite
radial distance in a star. If the pressure were a constant for
all proportions, then the mixed phase would be squeezed out of the
star because the pressure varies monotonically, being greatest at the
center and zero at the edge of the star.
In early work on phase transitions in neutron stars, they
were always treated as having constant pressure
at the zero temperature of the star, so the mixed phase never
appeared in any of those models.

The difference in properties of first order phase transitions
in a one-component substance and one with several independent
components or conserved charges is easily proven.
First consider a one-component substance for which equilibrium
of two phases, $A$ and $B$ is expressed by
\begin{eqnarray}
p_A(\mu, T)  = p_B(\mu, T)\,.
\end{eqnarray}
At constant $T$, 
the solution for the chemical potential is obviously unique and
independent of the proportion of the phases. So all properties
of the two phases remain unaltered as long as they are
in equilibrium, no matter the proportion.

Now consider a substance having two conserved charges 
(or
independent components). For definiteness we consider the
two independent conserved charges of neutron star matter,
baryon number and electric charge, whose densities we 
denote by
 $\rho$ and  $q$.
For the corresponding chemical potentials we choose
those of the neutron, $\mu_n$, and electron, $\mu_e$.
The chemical potentials of all other particles can be written
in terms of these independent ones.
Gibbs phase\index{Phase transition!phase equilibrium in multi-component substance}\index{Gibbs!phase equilibrium} 
equilibrium between the confined hadronic phase $C$ and 
the deconfined
quark matter\index{Quark matter} phase $D$
is now expressed as
\begin{eqnarray}
p_C(\mu_n,\mu_e,T) = p_D(\mu_n,\mu_e,T) \,.
\label{p}
\end{eqnarray}
This equation is insufficient to find the chemical potentials.
At fixed $T$,
it must be supplemented by another, say a statement of the
conservation\index{Conserved charge}\index{Neutron star!conserved charges} of one of the conserved charges. How should 
that statement be made? If one of the charges is the electric
charge, demanding that the electric charge density should vanish
identically in both phases
would satisfy the condition of charge neutrality\index{Charge neutrality}
as required of a star.\footnote{The
Coulomb force is so much stronger than
the gravitational that the net charge per baryon has to be less
than $10^{-36}$ which we can call zero.} That is in fact how charge 
neutrality in neutron stars was enforced for many years. However,
it is overly restrictive. The net charge must vanish, but the
charge density need not. Only the integrated electric charge density
must vanish, $\int q(r) d^3 r =0$.\footnote{Familiar examples
of neutral systems that have finite 
charge densities of opposite sign
are atoms and neutrons.} We refer to this as global
conservation\index{Phase transition!global conservation law} 
rather than local. It releases a degree of freedom 
that the physical system can exploit to find the minimum energy.
To express this explicitly, we note that according to the preparation
of the system, whether in the laboratory, or in a supernova, 
the concentration of the charges is fixed when the system is in a single
phase. Denote the concentration by
\begin{eqnarray}
c=Q/B \,.
\end{eqnarray}
However, when conditions of temperature or pressure
change to bring the system
into  two phases equilibrium, the concentrations
in each
can be different
\begin{eqnarray}
c_C=Q_C/B_C,~~~~~c_D=Q_D/B_D
\end{eqnarray}
provided only that the total charges are conserved,
\begin{eqnarray}
Q_C+Q_D=Q,~~~~~B_C+B_D=B\,.
\end{eqnarray}
Here $Q$ and $B$ denote the total
electric and baryon charge in a volume $V$.
The rearrangement of charges 
will take place to minimize the energy of the
system. The force that is responsible for exploiting this degree of 
freedom in neutron stars is the one responsible for the 
symmetry energy in nuclei and the valley of beta stability. 
Since neutron stars are far from symmetry, the symmetry energy
is quite large; the difference in Fermi energies of neutron and
proton, and the coupling of baryon isospin to the neutral rho 
meson are responsible.

For a uniform medium, and every sufficiently small 
region $V$ in a neutron 
star is uniform to high accuracy,
the statement of global conservation is\index{Conservation laws!for phases in equilibrium}
\begin{eqnarray}
\int_{V_C} q_C(r)\, d^3r + \int_{V_D} q_D(r)\, d^3r=
V_C\, q_C + V_D\, q_D = Q
\end{eqnarray}
where $V_C$ and $V_D$ denote the volume occupied by the two phases
respectively. This can be 
written more conveniently as
\begin{eqnarray}
(1-\chi)\, q_C(\mu_n,\mu_e) + \chi\, q_D(\mu_n,\mu_e)  
= Q/V \equiv \overline{q}
\label{c2}
\end{eqnarray}
where $q_C$ denotes the density of the conserved charge  in the
confined phase, $q_D$ in the deconfined phase,
 and 
\begin{eqnarray} 
\chi= V_D/V,~~~~V= V_C+V_D
\end{eqnarray}
is the volume proportion of phase $D$ and $\overline{q}$ is the
volume averaged electric charge (which for a star is zero). 
Now, equations (\ref{p}) and (\ref{c2})
are sufficient to find $\mu_n$ and $\mu_e$. But notice that
the solutions depend on the volume proportion $\chi$. Therefore,
also all properties of the two phases depend on their proportion,
including the common pressure. 
Having solved for the chemical potentials (and all field
quantities specified by their equations of motion),
the densities of the baryon conserved charge in the phases 
$C$ and $D$, are given by $\rho_C(\mu_n,\mu_e)$ and
$   \rho_D(\mu_n,\mu_e)$.  
The volume average of the baryon density
is given by an equation corresponding to Eq.\ (\ref{c2}).

Let us now discuss the consequences of opening the degree of freedom
embodied in Eq.\ (\ref{c2}), ie., in allowing electric charge (and
strangeness) to be
exchanged by the two phases in equilibrium so as to achieve the
minimum
energy at  the corresponding baryon density. Because of the 
long range of 
the Coulomb force
the Coulomb energy will be minimized when regions of like 
charge
are small, whereas the surface  interface energy will be minimized
when the surface areas of the regions of the two phases is small.
These\index{Neutron star!crystalline lattice}
 are opposing tendencies, and in first order can be reconciled
by minimizing their sum\index{Crystalline lattice}.
A Coulomb\index{Phase transition!Coulomb lattice} lattice will form
\cite{glen91:phase} of such a size and geometry of the
rare\index{Coulomb lattice} phase immersed at 
spacings\index{Phase transition!geometrical structure in mixed phase} \cite{glen95:c,glen95:e} 
in the dominant phase 
so as to 
minimize the 
energy. 

In better approximation, 
the total energy, consisting of the sum of bulk
energies, the surface and Coulomb energies and higher corrections
such as the curvature energy,
\begin{eqnarray}
E_{{\rm Total}} \approx E_{{\rm Bulk}} + E_{{\rm Surface}}
+ E_{{\rm Coulomb}} + E_{{\rm Curvature}} +  \cdots \,,
\label{total}
\end{eqnarray}
should be minimized. In still better approximation, the convenient
partition of the energy as above would be replaced by a
lattice  calculation of 
the total energy \cite{koonin85:a}.
In general, 
the opening of the degree(s) of freedom to conserve charges
globally rather than locally, or any other arbitrary way, can
only lower the total energy from the value it would have were the
degree of freedom closed, or in very special cases
leave it unchanged.  However, in the case of a neutron star,
the degree of freedom allows the bulk energy in the normal phase
to be lowered by decreasing the charge asymmetry of neutron star 
matter, so exchange of charge between the confined and deconfined 
phases is evidently favorable. It is unphysical to choose an
arbitrary
value of surface tension\index{Surface tension} such that the mixed phase is energetically
unfavored. The surface tension
 should be {\sl calculated
self-consistently}
by minimizing the total energy
Eq.\ (\ref{total}), when possible \cite{glen99:b}.  

\begin{figure}[t]
\vspace{-.24in}
\begin{center}
\includegraphics[width=.5\textwidth]{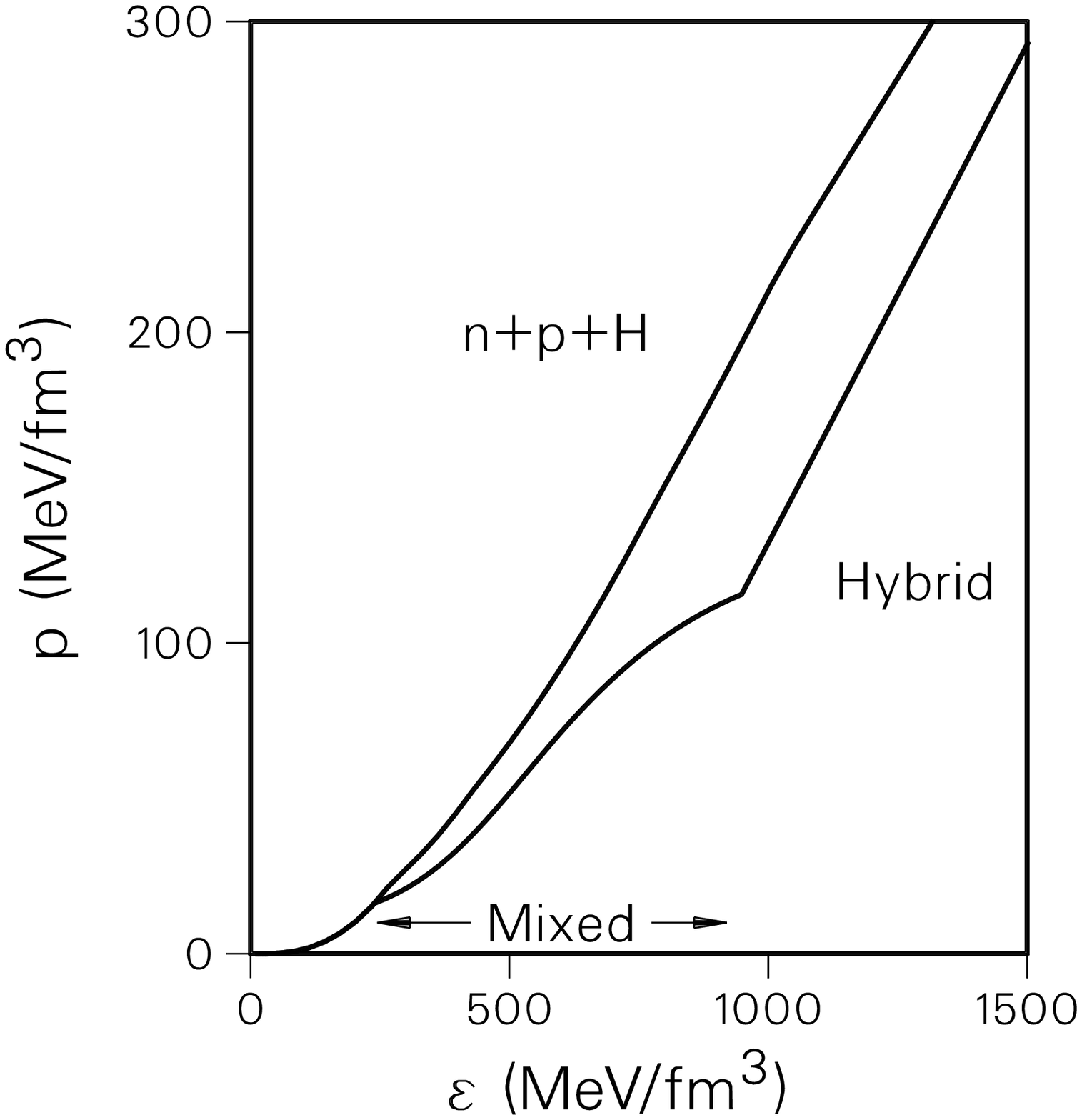}
\end{center}
\caption[]{Equation of state\index{Equation of state} 
 for\index{Equation of state!first order phase transition}
 a first order phase transition 
from neutron star matter in its confined to deconfined phases
(marked hybrid).\index{Neutron star!hybrid}
Note the monotonically increasing pressure. Comparison is made with
the\index{Equation of state}
 equation of state of neutron star matter in the confined phase
with nucleons\index{Hyperons!in neutron star} hyperons and leptons in equilibrium.} \label{h5}
\end{figure}

However, our purpose here is not to calculate the geometric
structure\footnote{See Ref.\
\cite{glen00:c} for a review.}
 of the mixed phase but to exhibit the equation of state
for\index{Phase transition!deconfinement} a first order deconfinement transition in neutron star matter
according to the above principles. Fig.\ \ref{h5} shows the
equation of state~ with the pure phases at low and high density and the mixed phase\index{Equation of state}
between. Note, as pointed out before that though the phase transition 
is first order, having as it does a  mixed phase, the pressure
(and all other internal properties) 
vary with density. The monotonic 
increase of pressure contrasts with early treatments of phase 
transitions in neutron stars prior to 1991, for which approximations 
rendered a constant pressure in the mixed phase
\cite{glen91:a}.   For a
 constant pressure phase transition, gravity will squeeze out the
mixed phase. Pressure is a monotonically decreasing function of 
distance
from the center of a star, just as it is in our atmosphere.

\subsection{Description of the Confined and Deconfined Phases}

To describe the confined phase of neutron star matter, we use
a\index{Relativistic mean-field theory} 
 generalization \cite{glen85:b}
   of relativistic nuclear field theory solved at the
mean field level in which 
nucleons and hyperons (the baryon octet)
are coupled to scalar, vector and vector-isovector mesons.
A full description of how the theory can be solved
for neutron star matter can be found in
\cite{books,glen85:b}.  
The Lagrangian is 
\begin{eqnarray}
{\cal L} & = &
\sum_{B} \overline{\psi}_{B} (i\gamma_{\mu} \partial^{\mu} - m_{B}
+g_{\sigma B} \sigma  - g_{\omega B} \gamma_{\mu} \omega^{\mu}
\nonumber \\[0ex]
& &
 - \frac{1}{2}       
 g_{\rho B} \gamma_{\mu} {\bf \tau} \cdot {\bf \rho}^{\mu} )
 {\psi}_{B}
 +\: \frac{1}{2}(\partial_{\mu} \sigma
  \partial^{\mu} \sigma - m_{\sigma}^{2} \sigma^{2})
 \nonumber \\[2ex]
 & &
   - \: \frac{1}{4} \omega_{\mu \nu}
  \omega^{\mu \nu} +\frac{1}{2} m_{\omega}^{2} \omega_{\mu} 
\omega^{\mu} 
     - \: \frac{1}{4}{\bf \rho}_{\mu \nu}\! 
\cdot\! {\bf \rho}^{\mu \nu}
  + \frac{1}{2} m_{\rho}^{2}{\bf \rho}_{\mu}\! \cdot\! 
{\bf \rho}^{\mu}  \nonumber \\[2ex]
     & &  
 -\; \frac{1}{3} b m_{n} (g_{\sigma} \sigma)^{3}
  - \frac{1}{4} c(g_{\sigma}\sigma)^{4}   
     \nonumber  \\[2ex]
      & &
       + \sum_{e^{-},\mu^{-}}
     \overline{\psi}_{\lambda} \bigl(i\gamma_{\mu}
       \partial^{\mu} - m_{\lambda} \bigr) \psi_{\lambda}\,.
	 \label{lagrangian}
	 \end{eqnarray}
The summ on $B$ is over all charege states of the baryon octet.
The parameters of the nuclear Lagrangian
can be algebaically determined  (see 2'nd ed. of \cite{books})
\index{Nuclear matter!saturation!properties}
 so that symmetric nuclear matter has the following
properties: binding energy of symmetric nuclear matter 
$B/A=-16.3$ MeV,
saturation density $\rho=0.153 {\rm ~fm^{-3}}$, compression modulus
$K=300$ MeV, symmetry energy coefficient $a_{{\rm sym}}=32.5$ MeV,
nucleon effective mass at saturation $m^{\star}_{{\rm sat}}=0.7m$ 
and
ratio of hyperon to nucleon couplings $x_\sigma=0.6,~x_\omega=0.653=x_\rho$
that yield, together with the foregoing parameters, the correct
$\Lambda$ binding in nuclear matter \cite{glen91:c}).

Quark matter\index{Quark matter}\index{MIT bag model}
is treated in a version of the MIT bag model with the three light
flavor quarks\index{Quark!light flavors}
($m_u=m_d=0,~m_s=150$ MeV)
as described in Ref.\ \cite{farhi84:a}.  A value of the
bag constant $B^{1/4}=180$ MeV is employed.\\[3ex]

{\bf Acknowledgments:} This work was supported by the
Director, Office of Energy Research,
Office of High Energy
and Nuclear Physics,
Division of Nuclear Physics,
of the U.S. Department of Energy under Contract
DE-AC03-76SF00098.~ F. Weber was supported by the Deutsche
Forschungsgemeinschaft.

\end{document}